\documentstyle[12pt,epsf,rotate]{article}
\input{psfig}
\newcommand{\gsim}{~{}_{\textstyle\sim}^{\textstyle >}~}

\def\OEE{\Omega_{\eta+\eta'}}
\def\as{\alpha_s}
\newcommand{\aem}{\alpha}

\def\klpn{K_{\rm L}\rightarrow\pi^0\nu\bar\nu}

\newcommand{\mz}{M_{\rm Z}}
\newcommand{\bea}{\begin{eqnarray}}
\newcommand{\eea}{\end{eqnarray}}
\newcommand{\bd}{\begin{displaymath}}
\newcommand{\ed}{\end{displaymath}}
\newcommand{\be}{\begin{equation}}
\newcommand{\ee}{\end{equation}}
\newcommand{\ord}{{\cal O}}

\newcommand{\eps}{\varepsilon}
\newcommand{\epe}{\varepsilon'/\varepsilon}
\newcommand{\mt}{m_{\rm t}}
\newcommand{\mc}{m_{\rm c}}
\newcommand{\ms}{m_{\rm s}}
\newcommand{\md}{m_{\rm d}}
\newcommand{\mw}{M_{\rm W}}
\newcommand{\gev}{\, {\rm GeV}}
\newcommand{\mev}{\, {\rm MeV}}
\newcommand{\bes}{B_6^{(1/2)}}
\newcommand{\bsi}{B_6^{(1/2)}}
\newcommand{\bei}{B_8^{(3/2)}}

\newcommand{\Lms}{\Lambda_{\overline{\rm MS}}}

\newcommand{\RE}{{\rm Re}}
\newcommand{\IM}{{\rm Im}}

\newcommand{\vcb}{|V_{cb}|}
\newcommand{\vtd}{|V_{td}|}
\newcommand{\vub}{|V_{ub}/V_{cb}|}

\begin{document}

% \bibliographystyle{physics}

%%%%%%%%%%%%%%%%%%%%%%%%%%%%%%%%%%%%%%%%%%%%%%%%%%%%%%%%%%%%%%%%%%%%%%%%%
% The titlepage
%%%%%%%%%%%%%%%%%%%%%%%%%%%%%%%%%%%%%%%%%%%%%%%%%%%%%%%%%%%%%%%%%%%%%%%%%
% \setcounter{footnote}{1}

\author{
{\large\bf S. Bosch${}^{1}$, A.J.~Buras${}^{1}$,
M. Gorbahn${}^{1}$, S. J\"ager${}^{1}$,} \\
{\large\bf 
 M. Jamin${}^{2}$,
M.E.~Lautenbacher${}^{1}$ and L. Silvestrini${}^{1}$} \\
\ \\
{\small\bf ${}^{1}$ Physik Department, 
Technische Universit\"at M\"unchen,} \\
{\small\bf D-85748 Garching, Germany} \\
{\small\bf ${}^{2}$ Sektion Physik, Universit\"at M\"unchen,} \\
{\small\bf Theresienstrasse 37, D-80333 M\"unchen, Germany}
}
\date{}
% \phantom{xxx} \vspace{-1.0cm}
\title{
{\normalsize\sf
\rightline{TUM-HEP-347/99}
\rightline{LMU 06/99}
}
% \bigskip
\bigskip
{\LARGE\bf
Standard Model Confronting\\  New Results for $\epe$
}}

\maketitle
\thispagestyle{empty}

\phantom{xxx} \vspace{-6mm}

\begin{abstract}
We analyze the CP violating ratio $\epe$ in the Standard Model in
view of the new KTeV results. We review the present status of the
most important non-perturbative parameters $\bsi$, $\bei$, $\hat B_K$
and of the strange quark mass $\ms$. 
We also briefly discuss the issues of final state interactions
and renormalization scheme dependence.
Updating the values of the CKM parameters,
of $\mt$ and $\Lms^{(4)}$ and using Gaussian errors for the experimental
input and flat distributions for the theoretical parameters we
find $\epe$ substantially below the NA31 and KTeV data:
$\epe= ( 7.7^{~+6.0}_{~-3.5}) \cdot 10^{-4}$ and
$\epe= ( 5.2^{~+4.6}_{~-2.7}) \cdot 10^{-4}$ in the NDR and HV
renormalization schemes respectively. A simple scanning of all
input parameters gives on the other hand
$1.05 \cdot 10^{-4} \le \epe \le 28.8 \cdot 10^{-4}$ and
$0.26 \cdot 10^{-4} \le \epe \le 22.0 \cdot 10^{-4}$ respectively.
Analyzing the dependence on various parameters we find that only 
for extreme values of $\bsi$, $\bei$ and $\ms$ as well as
suitable values of
CKM parameters and $\Lms^{(4)}$, the ratio $\epe$ can be made
consistent with data. We analyze the impact of these data on
the lower bounds for $\IM V_{td}V_{ts}^*$, $Br(K_L\to\pi^0\nu\bar\nu)$,
$Br(K_L\to\pi^0e^+e^-)_{\rm dir}$ and on $\tan\beta$ in
the Two Higgs Doublet Model II. 
 \end{abstract}

\newpage
\setcounter{page}{1}
\setcounter{footnote}{0}

%%%%%%%%%%%%%%%%%%%%%%%%%%%%%%%%%%%%%%%%%%%%%%%%%%%%%%%%%%%%%%%%%%%%%%%%%%%%%%%
% The main part of the paper
%%%%%%%%%%%%%%%%%%%%%%%%%%%%%%%%%%%%%%%%%%%%%%%%%%%%%%%%%%%%%%%%%%%%%%%%%%%%%%%
\section{Introduction}
\setcounter{equation}{0}

One of the most fascinating phenomena in particle physics is the violation of
CP symmetry in weak interactions. In the Standard Model CP violation is 
supposed to originate in a single complex phase $\delta$ in the charged 
current interactions of quarks \cite{KM}. 
This picture is consistent, within theoretical hadronic uncertainties, with 
CP violation in $K^0-\bar K^0$ mixing (indirect CP violation) discovered in 
$K_L\to\pi\pi$ decays already in 1964 \cite{CRONIN} and 
described by the parameter $\varepsilon$ \cite{PDG}:
\begin{equation} 
\varepsilon=(2.280\pm 0.013)\cdot 10^{-3} \exp(i \Phi_\varepsilon),
\qquad
 \Phi_\varepsilon\approx \frac{\pi}{4}.
\label{eexp}
\end{equation}
It is also consistent with the recent measurement of $\sin 2\beta$
from $B\to \psi K_S$ at CDF \cite{CDF99}, although the large experimental
error precludes any definite conclusion.

It should be emphasized that the agreement of 
the Standard Model with the experimental value of $\varepsilon$
is non-trivial as $|\sin\delta|\le 1 $.
Indeed in the Standard Model
\begin{equation}
\varepsilon= \hat B_K~ \IM\lambda_t \cdot 
             F_\eps(\mt,\RE\lambda_t) \exp(i \pi/4)
\label{epsm}
\end{equation}
where $\hat B_K$ is a non-pertubative 
parameter $\ord(1)$ and $\lambda_t=V_{td}V_{ts}^*$  
with $V_{ij}$ being the elements of the CKM matrix \cite{KM,CAB}. 
The function
$F_\eps$ results from well known box 
diagrams with $W^\pm,~ t,~c,~u$ exchanges \cite{IL} 
and includes NLO QCD corrections \cite{BJW90,HNab}. 
An explicit expression
for $F_\eps$ can be found in (10.42) of \cite{AJBLH}. 
It is an increasing function of 
the top quark mass 
$\mt$ and of $\RE\lambda_t$ . The QCD scale ($\Lms$) dependence of $F_\eps$ 
is very weak.

Now, $\IM\lambda_t$ 
is an important quantity as it plays a central role 
in the phenomenology of CP 
violation in $K$ decays and is furthermore closely related to the Jarlskog 
parameter $J_{CP}$ \cite{CJ}, 
the invariant measure of CP violation in the Standard Model: 
$J_{CP}=\lambda\sqrt{1-\lambda^2}\, \IM\lambda_t$ with $\lambda=0.221$
  denoting one of the Wolfenstein parameters \cite{WO}. 
To an excellent approximation one has
\be
\IM\lambda_t = |V_{ub}||V_{cb}|\sin\delta.
\label{imt}
\ee
As can be inferred from (\ref{epsm}) and (\ref{imt}) 
only for sufficiently
large values of $|V_{ub}|$, $|V_{cb}|$, $\mt$ and $\hat B_K$  
 can $\eps$ in (\ref{epsm}) and consequently the indirect CP violation in 
the Standard Model be consistent with the one observed experimentally.
It turns out that using the known values of 
$|V_{ub}|$, $|V_{cb}|$, $\mt$ (see Section 3)
 and taking $\hat B_K=0.80\pm 0.15$ in accordance with lattice 
and large-N calculations (see Section 2), 
the experimental value of $\eps$ can be reproduced 
in the Standard Model provided 
$\sin\delta \ge 0.69$. This determination of 
$\sin\delta$ includes constraints from $B^0_{d,s}-\bar B^0_{d,s}$ 
mixings. We also find
\be
1.04\cdot 10^{-4}\le\IM\lambda_t\le 1.63\cdot 10^{-4}~.
\label{imte}
\ee

It should be noticed that $\sin\delta=\ord(1)$ and that the extracted range 
for $\IM\lambda_t$ is not far from the 
upper limit of $1.73\cdot 10^{-4}$ following from the unitarity of the CKM 
matrix. It should also be 
emphasized that the large top quark mass plays an important role in 
obtaining the 
experimental value for $\eps$. 
Had $\mt$  been substantially lower than it is, 
the theoretical value of $\eps$ would be 
below the experimental one.

While indirect CP violation in $K_L\to\pi\pi$ reflects the fact that the 
mass eigenstates in the $K^0-\bar K^0$ system 
are not CP eigenstates, the so-called direct CP violation is realized 
via direct 
transitions between states of different CP parities: 
CP violation in the decay amplitude. 
In $K_L\to\pi\pi$ decays this type of CP violation is characterized by 
the parameter $\varepsilon'$. In the Standard Model one has
\be
\frac{\varepsilon'}{\varepsilon}=
\IM\lambda_t \cdot F_{\varepsilon'}(\mt,\Lms^{(4)},\ms,\bsi,\bei,\OEE)
\label{epeth}
\ee
where the function $F_{\varepsilon'}$ results from the calculation of 
QCD penguin and electroweak 
penguin diagrams. Here $\bsi$ and $\bei$ are non-perturbative parameters 
related to the dominant QCD penguin and electroweak penguin contributions 
respectively, $\Lms^{(4)}$ is the QCD scale and $\OEE$ represents isospin
breaking effects.

The expression (\ref{epeth}) has been obtained by calculating
$\varepsilon'$ and dividing it by the experimental value of
$\eps$ in (\ref{eexp}) in order to be able to compare
with the experimental value of $\epe$. 
This procedure exhibits the nature
of $\varepsilon'$ which representing direct CP violation is
proportional to $\IM\lambda_t$. However, one could also
proceed differently and ignoring the constraint (\ref{eexp})
calculate $\epe$ fully in theory. In this case (\ref{epeth})
is replaced by 
\be
\frac{\varepsilon'}{\varepsilon}=
\frac{\bar F_{\varepsilon'}(\mt,\Lms^{(4)},\ms,\bsi,\bei,\OEE)}
          {\hat B_K F_\eps(\mt,\RE\lambda_t)}
\label{epeth1}
\ee
where $\bar F_{\varepsilon'}=|\eps_{\rm exp}|F_{\varepsilon'}$
is independent of $\eps$. One should notice that $\IM\lambda_t$
cancelled out in  $\epe$ calculated in this manner and $\epe$
is actually a function of $\RE\lambda_t$ and not
of $\IM\lambda_t$. However, once the constraint (\ref{eexp})
has been taken into account (\ref{epeth1}) reduces to
(\ref{epeth}). We will return to this point in Section 3.

There is a long history of calculations of $\epe$ in the Standard
Model.
The first calculation of $\epe$ for $\mt \ll \mw$ without the inclusion
of renormalization group effects can be found in \cite{EGN}. 
Renormalization group effects
in the leading
logarithmic approximation have been first presented in \cite{GW79}. 
For $\mt \ll \mw$ only QCD
penguins play a substantial role. 
First extensive phenomenological analyses in this approximation
can be found in
\cite{BSS}.
Over the eighties these calculations
were refined through the inclusion of QED penguin effects 
for $\mt \ll \mw$ \cite{BW84,donoghueetal:86,burasgerard:87},
the inclusion of isospin breaking in the
quark masses \cite{donoghueetal:86,burasgerard:87,lusignoli:89},
and through improved estimates of hadronic matrix elements in
the framework of the $1/N$ approach \cite{bardeen:87}. 
This era of $\epe$ culminated
in the analyses in \cite{flynn:89,buchallaetal:90}, where QCD
penguins, electroweak penguins ($\gamma$ and $Z^0$ penguins)
and the relevant box diagrams were included for arbitrary
top quark masses. The strong cancellation between QCD penguins
and electroweak penguins for $m_t > 150~\gev$ found in these
papers was confirmed by other authors \cite{PW91}.

During the nineties considerable progress has been made by
calculating complete NLO corrections to $\varepsilon'$
\cite{BJLW1}-\cite{ROMA2}. Together with the NLO
corrections to $\varepsilon$ and $B^0-\bar B^0$ mixing
\cite{BJW90,HNab,Dresden}, this allowed
a complete NLO analysis of $\varepsilon'/\varepsilon$ including
constraints from the observed indirect CP violation ($\varepsilon$)
and  $B_{d,s}^0-\bar B_{d,s}^0$ mixings ($\Delta M_{d,s}$). The improved
determination of the $V_{ub}$ and $V_{cb}$ elements of the CKM matrix,
the improved estimates of hadronic matrix elements using the lattice 
approach as well as other non-perturbative approaches 
and in particular the determination of the top quark mass
$\mt$ had of course also an important impact on
$\varepsilon'/\varepsilon$. 

In a crude approximation (not to be used for any serious analysis)
\be\label{ap}
F_{\varepsilon'}\approx 13\cdot 
\left[\frac{110\mev}{\ms(2~\gev)}\right]^2
\left[\bsi(1-\OEE)-0.4\cdot \bei\left(\frac{\mt}{165\gev}\right)^{2.5}\right]
\left(\frac{\Lms^{(4)}}{340~\mev}\right)
\ee
where $\OEE\approx 0.25$. 
This formula exhibits very clearly the dominant uncertainties in 
$F_{\varepsilon'}$ which 
reside in the values of $\ms$, $\bsi$, $\bei$, $\Lms^{(4)}$ and $\OEE$. 
Because of the accurate value $\mt(\mt)=165\pm 5~\gev$, the uncertainty 
in $\epe$ due to the top quark mass amounts only to a few percent. 
A more accurate formula 
for $F_{\varepsilon'}$ will be given in Section 2.

A comparison of the formulae (\ref{epsm}) and (\ref{epeth}) reveals that
the analysis of $\eps$ is theoretically cleaner. Indeed, 
$\eps$ depends on a single
non-perturbative 
parameter $\hat B_K$, whereas $\epe$ is a sensitive function of $\bsi$,
$\bei$, $\ms$, $\Lms^{(4)}$ and $\OEE$. Moreover, the partial 
cancellation between QCD penguin ($\bsi$) and electroweak 
penguin ($\bei$) contributions requires 
accurate 
values of $\bsi$ and $\bei$ for an acceptable estimate of $\epe$.

Until recently the experimental situation on $\epe$ was rather unclear.
While the result of the NA31 collaboration at CERN with
Re$(\varepsilon'/\varepsilon) = (23.0 \pm 6.5)\cdot 10^{-4}$ \cite{barr:93}
clearly indicated direct CP violation, the value of E731 at Fermilab,
Re$(\varepsilon'/\varepsilon) = (7.4 \pm 5.9)\cdot 10^{-4}$
\cite{gibbons:93}, was compatible with superweak theories
\cite{wolfenstein:64} in which $\varepsilon'/\varepsilon = 0$.
This controversy is now settled with the very recent measurement
by KTeV at Fermilab \cite{KTEV}
\begin{equation}\label{eprime1}
\RE(\frac{\varepsilon'}{\varepsilon}) =
(28.0 \pm 4.1)\cdot 10^{-4} 
\end{equation}
which together with the NA31 result confidently establishes direct 
CP violation in nature.
The grand average including NA31, E731 and KTeV results reads
\be
\RE(\frac{\varepsilon'}{\varepsilon}) = (21.8\pm 3.0)\cdot 10^{-4}
\label{ga}
\ee
very close to the NA31 result but with a smaller error. The error 
should be further reduced once the first data from 
NA48 collaboration at
CERN are available and complete data from both collaborations have 
been analyzed. It is also 
of great interest to see what value for $\epe$ will be measured by KLOE at 
Frascati, which uses a different experimental technique than KTeV and NA48.

Does the direct CP violation observed in $K_L\to\pi\pi$ decays agree with 
the Standard Model expectations? Before entering the details let us take a 
set of ``central" values for the parameters entering $F_{\varepsilon'}$. 
Together with $\hat B_K=0.80$, $\mt(\mt)=165~\gev$, 
$|V_{ub}|=3.56\cdot 10^{-3}$ and
$|V_{cb}|=0.040$ needed for the $\eps$-analysis we set
\be
\bes=1.0, \qquad \bei=0.8, \qquad {\ms}(2~\gev)=110~\mev,
\qquad \OEE=0.25
\label{central}
\ee
and $\Lms^{(4)}=340~\mev$. 
Using the formula (\ref{eq:3b}) for  $F_{\varepsilon'}$, we find 
$F_{\varepsilon'}= 5.2$. On the other hand the $\eps$-analysis gives
$\IM \lambda_t=1.34\cdot 10^{-4}$. Consequently
\be
\left(\frac{\varepsilon'}{\varepsilon}\right)^{\rm central} = 
7.0 \cdot 10^{-4}
\label{cth}
\ee
well below the experimental findings in (\ref{ga}).

Equivalently, with $F_{\varepsilon'}=5.2 $, 
the experimental value in (\ref{ga})
implies $\IM\lambda_t=(4.2\pm0.6)\cdot 10^{-4} $ which lies outside the 
range (\ref{imte}) extracted from the standard analysis of the unitarity 
triangle. Moreover it 
violates the upper bound $\IM\lambda_t=1.73\cdot 10^{-4}$ following from 
the unitarity of the CKM matrix.

The fact that for central values of the 
input parameters the size of $\epe$
 in the Standard Model is well below the NA31 value of 
$(23.0\pm6.5)\cdot 10^{-4}$ has been known for some time. The 
extensive NLO analyses with lattice and large-N estimates of 
$\bes\approx 1$ and 
$\bei\approx 1$ performed 
first in \cite{BJLW,ROMA1} and after the top discovery in
\cite{ciuchini:95}-\cite{ciuchini:96} have found $\epe$ in 
the ball park of 
$(3-7)\cdot 10^{-4}$ for $\ms(2~\gev)\approx 130~\mev$. 
 On the other hand 
it has been stressed repeatedly in \cite{AJBLH,BJL96a}  that 
for extreme values of $\bes$, $\bei$ and $\ms$ still consistent with 
lattice, QCD sum rules and large-N estimates as well as 
sufficiently high values of $\IM\lambda_t$ and $\Lms^{(4)}$, 
a ratio $\epe$ as high as $(2-3)\cdot 10^{-3}$ could be obtained within 
the Standard Model. Yet, it has also been admitted that such simultaneously 
extreme values of all input parameters and consequently values of $\epe$
 close to the NA31 result 
are rather improbable in the Standard Model. 
Different conclusions have been reached in \cite{paschos:96}, where
values  $(1-2)\cdot 10^{-3}$ for $\epe$ can be found.
Also the Trieste group \cite{BERT98}, which calculated the parameters 
$\bes$ and $\bei$
 in the chiral quark model, found $\epe=(1.7\pm 1.4)\cdot 10^{-3}$.
On the other hand using an  effective chiral
lagrangian approach, the authors in \cite{BELKOV} found $\epe$
consistent with zero.

The purpose of the present paper is to update the analyses in 
\cite{AJBLH,BJL96a} and to 
confront the Standard Model estimates of $\epe$ with the experimental 
findings in (\ref{ga}).
Other very recent discussions of $\epe$ can be found in 
\cite{Nierste}-\cite{MM99}. We will comment on them
below.
In the present paper
we address in particular the following questions:

\begin{itemize}
\item
What is the maximal value of $\epe$ in the Standard Model consistent with 
the usual analysis of the unitarity triangle as a function of 
$\bes$, $\bei$, $\ms$ and $\Lms^{(4)}$ ?
\item
What is the lowest value of $\bes$ as a function of $\bei$ for fixed values 
of $\ms$ and $\Lms^{(4)}$ for which 
the Standard Model is simultaneously compatible with (\ref{ga})
 and the analysis of the unitarity triangle?
\item
What is the sensitivity of the analysis of $\epe$ to the values of 
$\OEE$ and $\hat B_K$?
\item
What is the impact of the experimental value for $\epe$ on
$\IM\lambda_t$, on the usual 
analysis of the unitarity triangle and in particular on Standard Model 
expectations for the rare decays 
$K_L\to\pi^0\nu\bar\nu$ and $K_L\to\pi^0e^+e^-$ in which direct 
CP violation plays an important role?
\item
What are the general implications of (\ref{ga}) for  physics beyond the 
Standard Model? In particular, what is the impact on the allowed range in 
the space $({\rm M_H}, \tan\beta)$ in the so called two Higgs doublet model
II (2HDMII) \cite{Abbott}?
\end{itemize}

While addressing these questions we would like to emphasize that it is by no 
means the purpose of our paper to fit 
$\bes$, $\bei$, $\ms$, $\Lms^{(4)}$, $\OEE$ and $\hat B_K$ in order to 
make the 
Standard Model compatible simultaneously with experimental values
on $\epe$, $\eps$ and the analysis of the unitarity triangle. Such an 
approach would be against the whole philosophy of searching for new physics 
with the help of loop induced transitions as represented by $\epe$ and 
$\eps$. Moreover it should be kept in mind that:

\begin{itemize}
\item
$\bes$, $\bei$ and $\hat B_K$, in spite of carrying the names of 
non-perturbative parameters, are really not 
parameters of the Standard Model as they can be calculated by means of 
non-perturbative methods in QCD. The same applies to $\OEE$.
\item
$\ms$, $\Lms^{(4)}$, $\mt$, $\vcb$ and $|V_{ub}|$ are parameters of the 
Standard Model but there are better places than $\epe$ to 
determine them. In particular the usual determinations of these parameters 
can only marginally be affected by  physics beyond the Standard Model,
which is   not necessarily the case for $\eps$ and $\epe$.
\end{itemize}

Consequently, the only parameter to be fitted by direct CP violation is 
$\sin\delta$ or $\IM\lambda_t$. The 
numerical analysis of $\epe$ as a function of 
$\bes$, $\bei$, $\ms$, $\Lms^{(4)}$, $\OEE$ and $\hat B_K$
should only give a global picture for 
which ranges of parameters the presence of new physics in $\epe$ and 
$\eps$ should be expected.

Our paper is organized as follows. In Section 2 we recall briefly 
the basic formulae for 
$\epe$ in the Standard Model. 
We also review the existing methods for estimating hadronic
matrix elements of relevant local operators and we present
a rather accurate analytic formula for $F_{\varepsilon'}$.
In Section 3 we address 
several of the questions listed 
above. 
In 
Section 4 we discuss briefly general implications for physics beyond 
the Standard Model. In particular we investigate the lower bound on
$\tan\beta$  as a function of the charged 
Higgs mass in the 2HDMII. 
Conclusions and outlook are given in 
Section 5.

\section{Basic Formulae}
\setcounter{equation}{0}
\subsection{Formulae for $\epe$}
The parameter $\varepsilon'$ is given in terms of the isospin amplitudes
$A_I$ as follows
\be\label{first}
\varepsilon'=\frac{1}{\sqrt{2}}\IM\left(\frac{A_2}{A_0}\right)
              \exp(i\Phi_{\varepsilon'}), 
   \qquad \Phi_{\varepsilon'}=\frac{\pi}{2}+\delta_2-\delta_0, 
\ee
where $\delta_I$ are final state interaction phases.
Then, the basic formula for $\epe$ is given by
\begin{equation}
\frac{\varepsilon'}{\varepsilon} = 
\IM \lambda_t\cdot F_{\varepsilon'},
\label{eq:epe1}
\end{equation}
where
\begin{equation}
F_{\varepsilon'} = 
\left[ P^{(1/2)} - P^{(3/2)} \right] \exp(i\Phi),
\label{eq:epe2}
\end{equation}
with
\begin{eqnarray}
P^{(1/2)} & = & r \sum y_i \langle Q_i\rangle_0
(1-\Omega_{\eta+\eta'})~,
\label{eq:P12} \\
P^{(3/2)} & = &\frac{r}{\omega}
\sum y_i \langle Q_i\rangle_2~.~~~~~~
\label{eq:P32}
\end{eqnarray}
Here
\begin{equation}
r = \frac{G_{\rm F} \omega}{2 |\eps| \RE A_0}~, 
\qquad
\langle Q_i\rangle_I \equiv \langle (\pi\pi)_I | Q_i | K \rangle~,
\qquad
\omega = \frac{\RE A_2}{\RE A_0}.
\label{eq:repe}
\end{equation}
Since
\begin{equation}
\Phi=\Phi_{\varepsilon'}-\Phi_\varepsilon \approx 0,
\label{Phi}
\end{equation}
$F_{\varepsilon'}$ and $\epe$
are real  to an excellent approximation.

The operators $Q_i$ are 
given explicitly  as follows:

{\bf Current--Current :}
\begin{equation}\label{OS1} 
Q_1 = (\bar s_{\alpha} u_{\beta})_{V-A}\;(\bar u_{\beta} d_{\alpha})_{V-A}
~~~~~~Q_2 = (\bar s u)_{V-A}\;(\bar u d)_{V-A} 
\end{equation}

{\bf QCD--Penguins :}
\begin{equation}\label{OS2}
Q_3 = (\bar s d)_{V-A}\sum_{q=u,d,s}(\bar qq)_{V-A}~~~~~~   
 Q_4 = (\bar s_{\alpha} d_{\beta})_{V-A}\sum_{q=u,d,s}(\bar q_{\beta} 
       q_{\alpha})_{V-A} 
\end{equation}
\begin{equation}\label{OS3}
 Q_5 = (\bar s d)_{V-A} \sum_{q=u,d,s}(\bar qq)_{V+A}~~~~~  
 Q_6 = (\bar s_{\alpha} d_{\beta})_{V-A}\sum_{q=u,d,s}
       (\bar q_{\beta} q_{\alpha})_{V+A} 
\end{equation}

{\bf Electroweak--Penguins :}
\begin{equation}\label{OS4} 
Q_7 = {3\over 2}\;(\bar s d)_{V-A}\sum_{q=u,d,s}e_q\;(\bar qq)_{V+A} 
~~~~~ Q_8 = {3\over2}\;(\bar s_{\alpha} d_{\beta})_{V-A}\sum_{q=u,d,s}e_q
        (\bar q_{\beta} q_{\alpha})_{V+A}
\end{equation}
\begin{equation}\label{OS5} 
 Q_9 = {3\over 2}\;(\bar s d)_{V-A}\sum_{q=u,d,s}e_q(\bar q q)_{V-A}
~~~~~Q_{10} ={3\over 2}\;
(\bar s_{\alpha} d_{\beta})_{V-A}\sum_{q=u,d,s}e_q\;
       (\bar q_{\beta}q_{\alpha})_{V-A} \,.
\end{equation}
Here, $\alpha,\beta$ are colour indices and
 $e_q$ denotes the electric quark charges reflecting the
electroweak origin of $Q_7,\ldots,Q_{10}$. 

The Wilson coefficient functions $ y_i(\mu)$
were calculated including
the complete next-to-leading order (NLO) corrections in
\cite{BJLW1}-\cite{ROMA2}. The details
of these calculations can be found there and in the review
\cite{BBL}. 
Their numerical values for $\Lms^{(4)}$ corresponding to
$\alpha_{\overline{MS}}^{(5)}(\mz)=0.119\pm 0.003$
and two renormalization schemes (NDR and HV)
are given in table 
\ref{tab:wc10smu13}.
There we also give the  coefficients $z_{1,2}$
relevant for the discussion of hadronic matrix elements.

\begin{table}[htb]
\caption[]{$\Delta S=1 $ Wilson coefficients at $\mu=\mc=1.3\gev$ for
$\mt=165\gev$ and $f=3$ effective flavours.
$y_1 = y_2 \equiv 0$.
\label{tab:wc10smu13}}
\begin{center}
\begin{tabular}{|c|c|c||c|c||c|c|}
\hline
& \multicolumn{2}{c||}{$\Lms^{(4)}=290\mev$} &
  \multicolumn{2}{c||}{$\Lms^{(4)}=340\mev$} &
  \multicolumn{2}{c| }{$\Lms^{(4)}=390\mev$} \\
\hline
Scheme & NDR & HV & NDR & HV & NDR & HV \\
\hline
$z_1$ & --0.393 & --0.477 & --0.425 & --0.521 & --0.458 & --0.570 \\
$z_2$ & 1.201 & 1.256 & 1.222 & 1.286 & 1.244 & 1.320 \\
\hline
$y_3$ & 0.027 & 0.030 & 0.030 & 0.034 & 0.033 & 0.038 \\
$y_4$ & --0.054 & --0.056 & --0.059 & --0.061 & --0.064 & --0.067 \\
$y_5$ & 0.006 & 0.015 & 0.005 & 0.016 & 0.003 & 0.017 \\
$y_6$ & --0.082 & --0.074 & --0.092 & --0.083 & --0.105 & --0.093 \\
\hline
$y_7/\aem$ & --0.038 & --0.037 & --0.037 & --0.036 & --0.037 & --0.034 \\
$y_8/\aem$ & 0.118 & 0.127 & 0.134 & 0.143 & 0.152 & 0.161 \\
$y_9/\aem$ & --1.410 & --1.410 & --1.437 & --1.437 & --1.466 & --1.466 \\
$y_{10}/\aem$ & 0.496 & 0.502 & 0.539 & 0.546 & 0.585 & 0.593 \\
\hline
\end{tabular}
\end{center}
\end{table}

It is customary in phenomenological
applications to take $\RE A_0$ and $\omega$ from
experiment, i.e.
\begin{equation}
\RE A_0 = 3.33 \cdot 10^{-7}\gev,
\qquad
\omega = 0.045,
\label{eq:ReA0data}
\end{equation}
where the last relation reflects the so-called $\Delta I=1/2$ rule.
This strategy avoids to a large extent the hadronic uncertainties 
in the real parts of the isospin amplitudes $A_I$.
In order to be consistent the constraint (\ref{eq:ReA0data})
should also be incorporated in the matrix elements $\langle Q_i\rangle_I$
necessary for the evaluation of $\epe$. This  in fact has 
has been done in \cite{BJLW} and we will return
to this approach briefly below. 
Studies of the $\Delta I=1/2$ rule can be found in
\cite{DI12,kilcup:99,DORT99}.

The sum in (\ref{eq:P12}) and (\ref{eq:P32}) runs over all contributing
operators. $P^{(3/2)}$ is fully dominated by electroweak penguin
contributions. $P^{(1/2)}$ on the other hand is governed by QCD penguin
contributions which are suppressed by isospin breaking in the quark
masses ($m_u \not= m_d$). The latter effect is described by

\begin{equation}
\Omega_{\eta+\eta'} = \frac{1}{\omega} \frac{(\IM A_2)_{\rm
I.B.}}{\IM A_0}\,.
\label{eq:Omegaeta}
\end{equation}
For $\Omega_{\eta+\eta'}$ we will first set
\begin{equation}
\Omega_{\eta+\eta'} = 0.25\,,
\label{eq:Omegaetadata}
\end{equation}
which is in the ball park of the values obtained in the $1/N$ approach
\cite{burasgerard:87} and in chiral perturbation theory
\cite{donoghueetal:86,lusignoli:89}. $\Omega_{\eta+\eta'}$ is
independent of $\mt$. We will investigate the sensitivity of $\epe$
to $\OEE$ in Section 3.

\subsection{Hadronic Matrix Elements}
The main source of uncertainty in the calculation of
$\epe$ are the hadronic matrix elements $\langle Q_i \rangle_I$.
They generally depend
on the renormalization scale $\mu$ and on the scheme used to
renormalize the operators $Q_i$. These two dependences are canceled by
those present in the Wilson coefficients $y_i(\mu)$ so that the
resulting physical $\epe$ does not (in principle) depend on $\mu$ and on the
renormalization scheme of the operators.  Unfortunately, the accuracy of
the present non-perturbative methods used to evalutate $\langle Q_i
\rangle_I$  is not
sufficient to have the $\mu$ and scheme dependences of
$\langle Q_i \rangle_I$ fully under control. 
We believe that this situation will change once the lattice calculations
and QCD sum rule calculations improve.
A brief review of the existing methods 
including most recent developments will be given below.

In view of this situation it has been suggested in \cite{BJLW} to
determine as many matrix elements $\langle Q_i \rangle_I$ as possible
from the leading CP conserving $K \to \pi\pi$ decays, for which the
experimental data is summarized in (\ref{eq:ReA0data}). 
To this end it turned out to be very convenient to determine $\langle
Q_i \rangle_I$ in the three-flavour effective theory at a scale $\mu
\approx m_c$. With this choice of $\mu$ the operators $Q_{1,2}^c$,
being present only for $\mu>\mc$,
are integrated out and the contribution of penguin operators to 
$\RE A_I$ turns out to be very small. Unfortunately, since the charm mass
is not much larger than the scale $M_K$ of the process we are
studying, the matching procedure between the four- and three-flavour
effective theories contains an ambiguity related to the choice of
external momenta in the matching \cite{BJLW1,BJLW}.
Furthermore, as pointed out in \cite{ciuchini:95}, there is an ambiguity
due to the contribution of higher dimensional operators which are
unsuppressed for $\mu \approx m_c$. However, all these ambiguities are of
$O(\alpha_s)$ and one can easily verify that their
possible contribution to  $\RE A_I$ is at the level of a few percent 
at most.
Consequently, they have only a minor impact on our determination of $\langle
Q_i \rangle_I$ at $\mu=\mc$ from  $\RE A_I$.
Using the renormalization group evolution one
can then find $\langle Q_i \rangle_I$ at any other scale $\mu \not=
\mc$. The details of this procedure can be found in
\cite{BJLW}. 

As we will see below this method allows to determine only the matrix
elements of the $(V-A)\otimes(V-A)$ operators. 
For the central value of $\IM\lambda_t$
these operators give a negative contribution to $\epe$ 
of about $-2.5\cdot 10^{-4}$. This shows that these
operators are only relevant if  $\epe$ is below $1 \cdot 10^{-3}$.
Unfortunately the matrix elements of the dominant $(V-A)\otimes(V+A)$
operators cannot be  determined by the CP conserving data and
one has to use  non-perturbative methods to estimate them.

Before giving the results for $\langle Q_i\rangle_I$ 
in our approach we would like to emphasize why it is reasonable 
to extract hadronic parameters from $\RE A_I$, while this would
not be the case for $\IM A_I$, which govern $\epe$. The point is that
$\RE A_I$, in contrast to $\IM A_I$, are not expected to be affected
by new physics contributions.

It is customary to express the matrix elements
$\langle Q_i \rangle_I$ in terms of non-perturbative parameters
$B_i^{(1/2)}$ and $B_i^{(3/2)}$ as follows:
\begin{equation}
\langle Q_i \rangle_0 \equiv B_i^{(1/2)} \, \langle Q_i
\rangle_0^{\rm (vac)}\,,
\qquad
\langle Q_i\rangle_2 \equiv B_i^{(3/2)} \, \langle Q_i
\rangle_2^{\rm (vac)} \,.
\label{eq:1}
\end{equation}
The label ``vac'' stands for the vacuum
insertion estimate of the hadronic matrix elements in question 
for
which $B_i^{(1/2)}=B_i^{(3/2)}=1$.

Then the approach in \cite{BJLW} gives at $\mu=\mc$:
\begin{eqnarray}
\langle Q_1(\mc) \rangle_0 &=&  \frac{0.187\gev^{3}}{z_1(\mc)}
 - \frac{z_2(\mc)}{z_1(\mc)} \langle Q_2(\mc) \rangle_0\, ,
\label{eq:Q10} \\
\langle Q_2(\mc) \rangle_0 &=&  \frac{5}{9} X B_2^{(1/2)} (\mc) \, ,
\label{eq:Q20} \\
\langle Q_3(\mc) \rangle_0 &=&  \frac{1}{3} X B_3^{(1/2)} (\mc) \, ,
\label{eq:Q30} \\
\langle Q_4(\mc) \rangle_0 &=&  \langle Q_3(\mc)\rangle_0 
    + \langle Q_2 (\mc) \rangle_0  -\langle Q_1 (\mc) \rangle_0 \, ,
\label{eq:Q40} \\
\langle Q_5 (\mc) \rangle_0 &=&  \frac{1}{3} B_5^{(1/2)}(\mc) 
                           \langle \overline{Q_6(\mc)} \rangle_0 \, ,
\label{eq:Q50} \\
\langle Q_6 (\mc) \rangle_0 &=&  -\,4 \sqrt{\frac{3}{2}} 
\left[ \frac{m_{\rm K}^2}{\ms(\mc) + \md(\mc)}\right]^2
\frac{F_\pi}{\kappa} \,B_6^{(1/2)}(\mc) \, ,
\label{eq:Q60} \\
\langle Q_7 (\mc) \rangle_0 &=& 
- \left[ \frac{1}{6} \langle \overline{Q_6(\mc)} \rangle_0 (\kappa + 1) 
         - \frac{X}{2} \right] B_7^{(1/2)} (\mc) \, ,
\label{eq:Q70} \\
\langle Q_8 (\mc) \rangle_0 &=& 
- \left[ \frac{1}{2} \langle \overline{Q_6(\mc)} \rangle_0 (\kappa + 1) 
         - \frac{X}{6} \right] B_8^{(1/2)} (\mc) \, ,
\label{eq:Q80} \\
\langle Q_9 (\mc) \rangle_0 &=& 
\frac{3}{2} \langle Q_1 (\mc) \rangle_0 
- \frac{1}{2} \langle Q_3(\mc) \rangle_0 \, ,
\label{eq:Q90} \\
\langle Q_{10}(\mc) \rangle_0 &=& 
    \langle Q_2(\mc) \rangle_0 + \frac{1}{2} \langle Q_1(\mc) \rangle_0
  - \frac{1}{2} \langle Q_3(\mc) \rangle_0 \, ,
\label{eq:Q100}
\end{eqnarray}
\begin{eqnarray}
\langle Q_1(\mc) \rangle_2 &=& 
\langle Q_2(\mc) \rangle_2 = \frac{8.44 \cdot 10^{-3}\gev^3}{z_+(m_c)} \, ,
\label{eq:Q122} \\
\langle Q_i \rangle_2 &=&  0 \, , \qquad i=3,\ldots,6 \, ,
\label{eq:Q362} \\
\langle Q_7(\mc) \rangle_2 &=& 
  -\left[ \frac{\kappa}{6 \sqrt{2}} \langle \overline{Q_6(\mc)} \rangle_0
          + \frac{X}{\sqrt{2}}
   \right] B_7^{(3/2)}(\mc) \, ,
\label{eq:Q72} \\
\langle Q_8(\mc) \rangle_2 &=& 
  -\left[ \frac{\kappa}{2 \sqrt{2}} \langle \overline{Q_6(\mc)} \rangle_0
          + \frac{\sqrt{2}}{6} X
   \right] B_8^{(3/2)}(\mc) \, ,
\label{eq:Q82} \\
\langle Q_9 (\mc) \rangle_2 &=& 
   \langle Q_{10}(\mc) \rangle_2 = 
\frac{3}{2} \langle Q_1(\mc) \rangle_2 \, ,
\label{eq:Q9102}
\end{eqnarray}
where
\begin{equation}
\kappa = \frac{F_\pi}{F_{\rm K} - F_\pi} \, ,
\qquad
X = \sqrt{\frac{3}{2}} F_\pi \left( m_{\rm K}^2 - m_\pi^2 \right) \, ,
\label{eq:XQi}
\end{equation}
and
\begin{equation}
\langle \overline{Q_6(\mc)} \rangle_0 =
   \frac{\langle Q_6(\mc) \rangle_0}{B_6^{(1/2)}(\mc)} \,,
\qquad z_+=z_1+z_2.
\label{eq:Q60bar}
\end{equation}
The equality of the matrix elements in (\ref{eq:Q122}) follows from 
isospin symmetry of strong interactions.
Finally,
 by making
 the very plausible assumption,
valid in known non-perturbative approaches, that
  $\langle Q_-(\mc) \rangle_0 \ge
\langle Q_+(\mc) \rangle_0 \ge 0$, where $Q_\pm=(Q_2\pm Q_1)/2$,
 $B_2^{(1/2)} (\mc)$ can be determined as well.
This gives  for $\Lms^{(4)}=340\mev$
\begin{equation}
B_{2,NDR}^{(1/2)}(\mc) =  6.5 \pm 1.0,
\qquad
B_{2,HV}^{(1/2)}(\mc) =  6.1 \pm 1.0 \, .
\label{eq:B122mc}
\end{equation}
The actual numerical values used for $m_{\rm K}$, $m_\pi$, $F_{\rm K}$,
$F_\pi$ are collected in the appendix of \cite{BBL}.
In particular $F_\pi=131~\mev$.

It should be noted that this method allows to determine not only
the size but also  the renormalization scheme
dependence of those matrix elements which can be fixed in this
manner. This dependence enters through $z_{1,2}(\mc)$ and the scheme
dependence of $B_2^{(1/2)} (\mc)$. In obtaining the results above
one also uses  operator relations valid for $\mu\le\mc$ which
allow to express $Q_4$, $Q_9$ and $Q_{10}$ in terms of $Q_1$, $Q_2$
and $Q_3$. Theoretical issues related to these relations in
the presence of NLO QCD corrections and the case of matrix elements
for $\mu>\mc$ are discussed in detail in \cite{BJLW}.

In order to proceed further one has to specify the remaining
$B_i$ parameters in the formulae above. 
As the numerical analysis in \cite{BJLW} shows $\epe$ is only
weakly sensitive to the values of the parameters
$B_3^{(1/2)}$, $B_5^{(1/2)}$, $B_7^{(1/2)}$, $B_8^{(1/2)}$
and $B_7^{(3/2)}$ as long as their absolute values are not
substantially larger than 1.
As in \cite{BJLW} our strategy
is to set
\begin{equation}
B_{3,7,8}^{(1/2)}(\mc) = 1,
\qquad
B_5^{(1/2)}(\mc) = B_6^{(1/2)}(\mc),
\qquad
B_7^{(3/2)}(\mc) = B_8^{(3/2)}(\mc)
\label{eq:B1278mc}
\end{equation}
and to treat $B_6^{(1/2)}(\mc)$ and $B_8^{(3/2)}(\mc)$ as free
parameters.

The approach in \cite{BJLW} allows then in a  good approximation
to express $\epe$ or equivalently $F_{\varepsilon'}$ in terms of
$\Lms^{(4)}$, $\mt$, $\ms$ and the two non-perturbative parameters 
$\bsi\equiv B_6^{(1/2)}(\mc)$ and $\bei\equiv B_8^{(3/2)}(\mc)$ 
which cannot be fixed by the CP conserving data.

\subsection{The Issue of Final State Interactions}
In (\ref{first}) and (\ref{Phi})
the strong phases $\delta_0\approx 37^\circ$ and
$\delta_2\approx -7^\circ$ are taken from experiment.
They can also be calculated from NLO chiral perturbation
for $\pi\pi$ scattering \cite{JGUM}. 
However, generally 
non-perturbative approaches to hadronic matrix elements are unable 
to reproduce them at present. 
As $\delta_I$ are factored out in (\ref{first}), in non-perturbative
calculations in which some final state interactions are
present in $\langle Q_i\rangle_I$ 
one should make the following
replacements in (\ref {eq:P12}) and (\ref{eq:P32}):
\be\label{FS0} 
\langle Q_i\rangle_I \to \frac{\RE\langle Q_i\rangle_I}
{(\cos\delta_I)_{\rm th}}
\ee
in order to avoid double counting of final state interaction
phases. Here $(\cos\delta_I)_{\rm th}$ is obtained in a given
non-perturbative calculation. In leading large-N calculations
and in quenched lattice calculations the phases $\delta_I$
vanish and this replacement is ineffective. When loop
corrections in the large-N approach \cite{bardeen:87,DORT99,DORT98}
and in the chiral quark model \cite{BERT98} are included
an absorptive part and related non-vanishing phases are
generated. Yet, in most calculations the phases are 
substantially smaller than found in experiment. For instance
in the chiral quark model $(\cos\delta_0)_{\rm th}\approx 0.94$
to be compared to the experimental
value $(\cos\delta_0)_{\rm exp}\approx 0.8$. Even smaller
phases are found in \cite{bardeen:87,DORT99,DORT98}.

The above point has been first discussed by the Trieste group
\cite{BERT98} who suggested
that in models in which at least the real part of $\langle Q_i\rangle_I$
can be calculated reliably, one should make the following
replacements in (\ref {eq:P12}) and (\ref{eq:P32}):
\be\label{FS} 
\langle Q_i\rangle_I \to \frac{\RE\langle Q_i\rangle_I}
{(\cos\delta_I)_{\rm exp}}
\ee
where this time the experimental value of $\delta_I$ enters the
denominator.
As $(\cos\delta_0)_{\rm exp}\approx 0.8$ and 
$(\cos\delta_2)_{\rm exp}\approx 1$
this modification enhances $P^{(1/2)}$ by $25\%$ leaving $P^{(3/2)}$
unchanged. The same procedure has been adopted in \cite{DORT99}. 
To our knowledge there is no method for hadronic matrix
elements which can provide $\delta_0\approx 37^\circ$  and consequently 
the replacement (\ref{FS}) may lead to an overestimate of the matrix 
elements.

As in our paper the matrix elements of $(V-A)\otimes(V-A)$
operators are extracted from the data, the replacements in
(\ref{FS0}) and (\ref{FS}) are ineffective for the
determination of the corresponding contributions. They
merely change the definition of the $B_i$ parameters
in the matrix elements of $(V-A)\otimes(V-A)$ operators.
The situation is different with the matrix elements
of $(V-A)\otimes(V+A)$ operators which are taken from
theory.
Yet in view of the remarks made above, in our analysis we will use
exclusively  (\ref {eq:P12}) and (\ref{eq:P32}) including
possible  effects of this sort in the uncertainties in
 $\bsi$ and $\bei$.

\subsection{An Analytic Formula for $\epe$}
           \label{subsec:epeanalytic}
As shown in \cite{buraslauten:93}, it is possible to cast the formal
expressions for $\epe$ in (\ref{eq:epe1})--(\ref{eq:P32})
into an analytic formula which exhibits the $\mt$ dependence
together with the dependence on $\ms$, $\Lms^{(4)}$,
$B_6^{(1/2)}$ and $B_8^{(3/2)}$. 
To this end the approach for hadronic matrix elements presented
above is used and $\OEE$ is set to $0.25$.
The analytic formula given below, while being rather accurate, 
exhibits
various features which are not transparent in a pure numerical
analysis. It can be used in phenomenological applications if
one is satisfied with a few percent accuracy. Needless to
say, in our numerical analysis in Section 3 we have used
exact expressions.

In this formulation
the function $F_{\varepsilon'}$
is given simply as follows ($x_t=\mt^2/\mw^2$):
\begin{equation}
F_{\varepsilon'} =
P_0 + P_X \, X_0(x_t) + P_Y \, Y_0(x_t) + P_Z \, Z_0(x_t) 
+ P_E \, E_0(x_t). 
\label{eq:3b}
\end{equation}
Exact expressions for the $\mt$-dependent functions in (\ref{eq:3b}) 
can be found for instance in \cite{AJBLH,BBL}. 
In the range $150\gev \le \mt \le 180\gev$ 
one has  to an accuracy much better than 1\% 
\begin{equation}
X_0(x_t)=1.51~\left(\frac{\mt}{165\gev}\right)^{1.13},
\quad\quad
Y_0(x_t)=0.96~\left(\frac{\mt}{165\gev}\right)^{1.55},
\end{equation}
\begin{equation}
 Z_0(x_t)=0.66~\left(\frac{\mt}{165\gev}\right)^{1.90},\quad\quad
   E_0(x_t)= 0.27~\left(\frac{\mt}{165\gev}\right)^{-1.08}.
\end{equation}
In our numerical analysis we use exact expressions.
 
The coefficients $P_i$ are given in terms of $B_6^{(1/2)} \equiv
B_6^{(1/2)}(\mc)$, $B_8^{(3/2)} \equiv B_8^{(3/2)}(\mc)$ and $\ms(\mc)$
as follows:
\begin{equation}
P_i = r_i^{(0)} + 
r_i^{(6)} R_6 + r_i^{(8)} R_8 \, .
\label{eq:pbePi}
\end{equation}
where
\be\label{RS}
R_6\equiv \bsi\left[ \frac{137\mev}{\ms(\mc)+\md(\mc)} \right]^2,
\qquad
R_8\equiv \bei\left[ \frac{137\mev}{\ms(\mc)+\md(\mc)} \right]^2.
\ee
The $P_i$ are renormalization scale and scheme independent. They depend,
however, on $\Lms^{(4)}$. In table~\ref{tab:pbendr} we give the numerical
values of $r_i^{(0)}$, $r_i^{(6)}$ and $r_i^{(8)}$ for different values
of $\Lms^{(4)}$ at $\mu=\mc$ in the NDR renormalization scheme. 
This table differs from the ones presented in \cite{AJBLH,BJL96a}
in the values of $\Lms^{(4)}$ and the central value of $\ms(\mc)$
in $R_s$ which
has been lowered from $150\mev$ to $130\mev$.
The
coefficients $r_i^{(0)}$, $r_i^{(6)}$ and $r_i^{(8)}$ depend only very
weakly on
$\ms(\mc)$ as the dominant $\ms$ dependence has been factored out. The
numbers given in table~\ref{tab:pbendr} correspond exactly 
to $\ms(\mc)=130\,\mev$.
However, even for $\ms(\mc)\approx100\mev$ or
$\ms(\mc)\approx160\mev$, the analytic expressions given
here reproduce the numerical calculations of $\epe$ given in Section 3 
to better than $4\%$.
For different scales $\mu$ the numerical values in the tables change
without modifying the values of the $P_i$'s as it should be. The values
of $B_6^{(1/2)}$ and $B_8^{(3/2)}$ should also be  modified, 
in principle, but as a detailed numerical analysis in \cite{BJLW}
showed, it is a good approximation to keep them $\mu$-independent
for $1~\gev \le\mu\le 2~\gev$. We will return to this point below.

\begin{table}[thb]
\caption[]{Coefficients in the formula (\ref{eq:pbePi})
 for various $\Lms^{(4)}$ in 
the NDR scheme.
The last row gives the $r_0$ coefficients in the HV scheme.
\label{tab:pbendr}}
\begin{center}
\begin{tabular}{|c||c|c|c||c|c|c||c|c|c|}
\hline
& \multicolumn{3}{c||}{$\Lms^{(4)}=290\mev$} &
  \multicolumn{3}{c||}{$\Lms^{(4)}=340\mev$} &
  \multicolumn{3}{c| }{$\Lms^{(4)}=390\mev$} \\
\hline
$i$ & $r_i^{(0)}$ & $r_i^{(6)}$ & $r_i^{(8)}$ &
      $r_i^{(0)}$ & $r_i^{(6)}$ & $r_i^{(8)}$ &
      $r_i^{(0)}$ & $r_i^{(6)}$ & $r_i^{(8)}$ \\
\hline
0 &
   --2.771 &   9.779 &   1.429 &
   --2.811 &  11.127 &   1.267 &
   --2.849 &  12.691 &   1.081 \\
$X_0$ &
     0.532 &   0.017 &       0 &
     0.518 &   0.021 &       0 &
     0.506 &   0.024 &       0 \\
$Y_0$ &
     0.396 &   0.072 &       0 &
     0.381 &   0.079 &       0 &
     0.367 &   0.087 &       0 \\
$Z_0$ &
     0.354 &  --0.013 &   --9.404 &
     0.409 &  --0.015 &  --10.230 &
     0.470 &  --0.017 &  --11.164 \\
$E_0$ &
     0.182 &  --1.144 &   0.411 &
     0.167 &  --1.254 &   0.461 &
     0.153 &  --1.375 &   0.517 \\
\hline
0 &
   --2.749 &   8.596 &   1.050 &
   --2.788 &   9.638 &   0.871 &
   --2.825 &  10.813 &   0.669 \\
\hline
\end{tabular}
\end{center}
\end{table}

The inspection of table~\ref{tab:pbendr} shows
that the terms involving $r_0^{(6)}$ and $r_Z^{(8)}$ dominate the ratio
$\epe$. Moreover, the function $Z_0(x_t)$ representing a gauge invariant
combination of $Z^0$- and $\gamma$-penguins grows rapidly with $\mt$
and due to $r_Z^{(8)} < 0$ these contributions suppress $\epe$ strongly
for large $\mt$ \cite{flynn:89,buchallaetal:90}.

\subsection{Renormalization Scheme Dependence}\label{rsd}
Concerning the renormalization scheme dependence only the coefficients
$r_0^{(0)}$, $r_0^{(6)}$ and $r_0^{(8)}$
are scheme dependent at the NLO level. Their values in the HV
scheme are given in the last row of table~\ref{tab:pbendr}.
We note that the parameter $r_0^{(0)}$ is essentially the same in both
schemes as the dominant scheme independent contributions to
$r_0^{(0)}$ have been determined by the data on $\RE A_I$. Since
$P_0$ must be scheme independent and $r_0^{(6)}$ and $r_0^{(8)}$
are scheme dependent, we conclude that $\bsi$ and $\bei$ must
be scheme dependent. Indeed the matrix elements in the NDR
and HV schemes are related by a finite renormalization
which can be found in equation (3.7) of \cite{BJLW}. Using this
equation together with the approach to matrix elements presented
above, we find  approximate relations between the values of 
$(\bsi,\bei)$ in 
the NDR scheme and the corresponding values in the HV scheme:
\be\label{NDRHV}
(\bsi)_{\rm HV}\approx 1.2 (\bsi)_{\rm NDR},
\qquad
 (\bei)_{\rm HV}\approx 1.2 (\bei)_{\rm NDR}.
\ee 
One can check that the scheme dependence 
of $(\bsi,\bei)$ cancels to a very good approximation the one
of $r_0^{(6)}$ and $r_0^{(8)}$ so that $P_0$ is scheme independent.

On the other hand the coefficients $r_i$, 
$i=X, Y, Z, E$ are  scheme independent at NLO. 
This is related to the fact that the $\mt$
dependence in $\epe$ enters first at the NLO level and consequently all
coefficients $r_i$ in front of the $\mt$ dependent functions must be
scheme independent. Strictly speaking then the scheme dependence
of $B_6^{(1/2)}$ and $B_8^{(3/2)}$ inserted into
$P_i$ with $i \not= 0$ is really a part of 
higher order contributions to $\epe$ and should be dropped
at the NLO level. Formally this can be done by not performing the
finite renormalization when going from the NDR to the HV scheme.
Then the coefficients $P_i$ with $i \not= 0$ are clearly scheme 
independent.

In practice the situation is more complicated. The present
non-perturbative methods used to evaluate $\bsi$ and $\bei$
like the large-N approach are not sensitive
to the renormalization scheme dependence and we do not know
 which renormalization scheme the resulting values for
these parameters correspond to. Lattice calculations,
QCD sum rule calculations and the chiral quark model 
can in principle give us the
scheme dependence of $\bsi$ and $\bei$ but the accuracy of
these methods must improve before they could
 be useful in this respect.

In view of this situation our strategy will be to use the
same values for $\bsi$ and $\bei$ in the NDR and HV schemes.
This will introduce a scheme dependence in $P_0$ and
consequently in $\epe$ but will teach us something about
the uncertainty in $\epe$ due to the poor sensitivity of present
methods to renormalization scheme dependence.

It should  also be noted that even if we knew the scheme dependence
of $\bsi$ and $\bei$ without the ability of separating a scheme
independent part in these parameters, the resulting $\epe$ would
be scheme dependent at the NLO level. This time 
the scheme dependence would enter through the scheme dependence of
$P_i$ with $i \not= 0$. The latter scheme dependence could
only be reduced by including the next order of perturbation theory
in the Wilson coefficients:
a formidable task. We should also stress \cite{BJLW}
that the scheme dependences discussed here apply not only
to QCD corrections but also to QED corrections. That is
QED corrections to the matrix elements of operators have to be also
known.

For similar reasons the NLO analysis of $\epe$ is still insensitive to
the precise definition of $\mt$. In view of the fact that the NLO
calculations needed to extract $\IM \lambda_t$  
have been performed with $\mt=\mt(\mt)$ we will also use  this 
definition in calculating $F_{\varepsilon'}$.

\subsection{Status of the Strange Quark Mass}

At this point it seems appropriate to summarize the present status of the
value of the strange quark mass. Since different methods provide
$\ms$ at different values of $\mu$ we give in table~\ref{tab:ms}
 a dictionary between
the $\ms$ values at $\mu=1\gev$, $\mu=\mc=1.3\gev$ and $\mu=2\gev$.

 In the case of quenched lattice QCD the present status
has been summarized recently by Kenway \cite{kenway98}. 
Averaging the results
presented by him at LATTICE 98, we obtain $\ms(2\gev)=(120\pm20)\,\mev$. It
is expected that unquenching will lower this value but it is difficult to
tell by how much. Strange quark masses as low as 
$\ms(2\gev)=80\,\mev$ have been
reported in the literature \cite{LOWMS}, 
although the errors on unquenched calculations
are still large. Lacking more precise information on unquenched lattice
calculations we take as the average lattice value 
\be\label{mslat}
\ms(2\gev)=(110\pm20)\;\mev.
\ee
which is very close to the one given by Gupta \cite{GUPTA98}.

A large number of determinations of the strange quark mass from QCD sum
rules exist in the literature. Historically, QCD sum rule results for $\ms$
are given at a scale $1\,\gev$. Taking an average over recent results
\cite{jaminmuenz:95}-\cite{dominguezetal:98}
we find $\ms(1\gev)=(170\pm30)\,\mev$. This translates to
$\ms(2\gev)=(124\pm22)\,\mev$, somewhat higher than the lattice result
but compatible within the errors. QCD sum rules also allow to
derive lower bounds on the strange quark mass. It was found that generally
$\ms(2\gev)\gsim 100\,\mev$ \cite{DERAF}-\cite{Dosch}. If these bounds
hold, they would rule out the very low strange mass values found in
unquenched lattice QCD simulations.

\begin{table}[thb]
\caption[]{The dictionary between the values of $m_s(\mu)$ 
in units of $\mev$. $\Lms^{(4)}=340~\mev$ and $\mc=1.3~\gev$
have been used.
\label{tab:ms}}
\begin{center}
\begin{tabular}{|c|c|c|c|c|}\hline
  $\ms(\mc)$& $ 105$& $130$& $155$ & $180$  \\ \hline
 $\ms(2~\gev)$& $ ~90$& $111$& $132$ & $154$ \\ \hline
 $\ms(1~\gev)$& $ 123$& $152$& $181$ & $211$ \\ \hline
  \end{tabular}
\end{center}
\end{table}

Finally, one should also mention the very recent determination of the
strange mass from the hadronic $\tau$-spectral
  function \cite{PP,aleph:99} which proceeds similarly
 to the determination of
$\alpha_s$ from $\tau$-decays. Normalized at the $\tau$ mass, 
the ALEPH collaboration obtains 
$\ms(m_\tau)=(176^{+46}_{-57})\,\mev$ which translates to
$\ms(2\gev)=(170^{+44}_{-55})\,\mev$. We observe that the central value
is much larger than the corresponding results from lattice and sum rules
although the error is still large. In the future, however, improved
experimental statistics and a better understanding of perturbative QCD
corrections should make the determination of $\ms$ from the $\tau$-spectral
function competitive to the other methods.

We conclude that the error on $\ms$ is still rather large. In our 
numerical analysis of $\epe$, where $\ms$ is evaluated at the scale $\mc$,
we will set
\begin{equation}\label{msvalues}
\ms(\mc)=(130\pm25)\;\mev \,,
\end{equation}
roughly corresponding to $\ms(2~\gev)$ given in (\ref{mslat}).

\subsection{Review of $\hat B_K$, 
 $B^{(1/2)}_{6}$ and $B^{(3/2)}_{8}$ }
\subsubsection{$\hat B_K$}
The renormalization group invariant parameter $\hat B_K$ is
defined through
\begin{equation}
\hat B_K = B_K(\mu) \left[ \alpha_s^{(3)}(\mu) \right]^{-2/9} \,
\left[ 1 + \frac{\alpha_s^{(3)}(\mu)}{4\pi} J_3 \right],
\label{eq:BKrenorm}
\end{equation}

\begin{equation}
\langle \bar K^0| (\bar s d)_{V-A} (\bar s d)_{V-A} |K^0\rangle
\equiv \frac{8}{3} B_K(\mu) F_K^2 m_K^2
\label{eq:KbarK}
\end{equation}
where $J_3=1.895$ and $J_3=0.562$ in the NDR and HV scheme 
respectively.

There is a long history of evaluating  $\hat B_K$
in various non-perturbative approaches. 
The status of quenched lattice calculations \cite{JLQCD,GKS,APE}
as of 1998 has been 
reviewed by Gupta \cite{GUPTA98}.
The most accurate result for $B_K(2~\gev)$
using lattice methods has been obtained by the 
JLQCD collaboration  \cite{JLQCD}:
$B_K(2~\gev)=0.628\pm0.042$. 
A similar result has been published by Gupta, Kilcup and
Sharpe \cite{GKS} last year.
The APE collaboration \cite{APE} 
found $B_K(2~\gev)=0.66\pm0.11$ which is
consistent with \cite{JLQCD,GKS}. 
The final  lattice value
given by Gupta was then
\be\label{G1}
(\hat B_K)_{\rm Lattice}=0.86\pm0.06\pm0.06
\ee
where the second error is attributed to quenching.
 The
corresponding result from the APE collaboration \cite{APE}
was $\hat B_K=0.93\pm0.16$. The most recent global analysis of
lattice data 
including also the UKQCD results gives \cite{LL}
\be
\hat B_K=0.89\pm0.13
\ee
in good agreement with (\ref{G1}).

In the $1/N$ approach of \cite{bardeen:87} one finds $\hat B_K=0.70\pm 0.10$
\cite{BBG0,Bijnens}. The most recent analysis in this approach
with a modified matching procedure and inclusion of higher order
terms in momenta gives a bigger range $0.4<\hat B_K<0.7 $ \cite{DORT99}
which results from a stronger dependence on the matching scale between
short and long distance contributions than found in previous 
calculations. It is hoped that inclusion of
higher resonances in the effective low energy theory will make
the dependence weaker.

QCD sum rules give results around $\hat B_K=0.5-0.6$ with
errors in the range $0.2-0.3$ \cite{BKQCD}.
Still lower
values are  found using the QCD Hadronic Duality
approach   ($\hat B_K=0.39\pm0.10$) \cite{Prades}, 
the SU(3) symmetry and
PCAC ($\hat B_K=1/3$) \cite{Donoghue} or chiral perturbation theory
at next-to-leading order ($\hat B_K =0.42\pm 0.06$) \cite{Bruno}.
However, as stressed in \cite{Bijnens,Sonoda},  SU(3) breaking
effects  considerably increase these values.
Finally, the analysis in
the chiral quark model gives a value as high as 
$\hat B_K=1.1\pm 0.2$ \cite{BERT97}. 

In our numerical analysis presented 
below we will use 
\begin{equation}\label{BKT}
\hat B_K=0.80\pm 0.15 \,
\end{equation}
which is in the ball park of various lattice and large-N estimates.
We will, however, discuss what happens if values outside this
range are used.
\subsubsection{General Comments on $\bsi$ and $\bei$}
As the different methods for the evaluation of these parameters
use different values of $\mu$, it is useful to say something about
their $\mu$-dependence. As seen in (\ref{eq:Q60}) and (\ref{eq:Q82})
the $\mu$-dependences of
$\langle Q_6 (\mu) \rangle_0$ and $\langle Q_8 (\mu) \rangle_2$
are governed by the known $\mu$-dependence of $\ms$ and $\md$
and could also in principle be present in $\bsi(\mu)$ and
$\bei(\mu)$. 

Now, as can be demonstrated  in the large-N limit,
the $\mu$-dependence
of $1/(\ms(\mu)+\md(\mu))^2$ in $\langle Q_6\rangle_0$ and 
$\langle Q_8\rangle_2$
is exactly cancelled in the decay amplitude by the diagonal
evolution (no operator mixing) of the Wilson coefficients $y_6(\mu)$ and
$y_8(\mu)$ taken in the large-N limit. An explicit demonstration 
of this feature is given in \cite{AJBLH}. In the large-N limit
one also finds
\be\label{LN}
B^{(1/2)}_{6}=B^{(3/2)}_{8}=1, \quad\quad{\rm (Large-N~Limit)}. 
\ee

The $\mu$-dependence of $\bsi$ and $\bei$ for $N=3$ and
in the presence of mixing with other operators
has been investigated in \cite{BJLW}. This analysis shows that
$\bsi$ and $\bei$ depend only very weakly on $\mu$,
when $\mu\ge 1~\gev$. In such a numerical
renormalization study the
factors $B_{6}^{(1/2)}$ and $B_{8}^{(3/2)}$  have been set to unity 
at $\mu=\mc$.
Subsequently the evolution of the matrix elements in the range $1\gev
\le \mu \le 2\gev$ has been calculated showing that for the NDR scheme
$B_{6}^{(1/2)}$ and $B_{8}^{(3/2)}$ were $\mu$ independent within
an accuracy of $2\,\%.$ The $\mu$ dependence in the HV scheme has
been found to be stronger but still below $6\,\%$.
Similar weak $\mu$-dependences have been found
for $B_{5}^{(1/2)}$ and $B_{7}^{(3/2)}$.

These findings simplify the comparison of results for
$B_{5,6}^{(1/2)}$ and $B_{7,8}^{(3/2)}$ obtained by different
methods.
\subsubsection{$B^{(1/2)}_{6}$ and $B^{(3/2)}_{8}$ from the Lattice}
The lattice calculations of $B_{5,6}^{(1/2)}$ and $B_{7,8}^{(3/2)}$
 have been reviewed
by Gupta \cite{GUPTA98} and the APE collaboration \cite{APE}. 
They are all given at $\mu=2\gev$ and in the NDR scheme.
The most reliable
results are found for $B^{(3/2)}_{7,8}$. The ``modern" quenched
estimates for these parameters
are collected in table \ref{tab:317} \cite{GUPTA98}.
The errors given there are purely statistical. 
The first three calculations
use perturbative matching between lattice and continuum, the last
one uses non-perturbative matching. All three groups agree
within perturbative matching that $B_{7,8}^{(3/2)}$ 
are suppressed below unity:
$B_{7}^{(3/2)}\approx 0.6$ and $B_{8}^{(3/2)}\approx 0.8$.
The non-perturbative matching
seems to increase these results by about $20\%$.
It is important to see whether this feature will be confirmed
by other groups.

Concerning the
lattice results for $B^{(1/2)}_{5,6}$ the situation is
 worse. The old results read
$B^{(1/2)}_{5,6}(2~\gev)=1.0 \pm 0.2$ \cite{kilcup:91,sharpe:91}.
More accurate estimates for $B^{(1/2)}_{6}$ have been given
in \cite{kilcup:98}: 
$B^{(1/2)}_{6}(2~\gev)=0.67 \pm 0.04\pm 0.05$
(quenched) and $B^{(1/2)}_{6}(2~\gev)=0.76 \pm 0.03\pm0.05$
($f=2$). However, as stressed by Gupta \cite{GUPTA98}, the systematic
errors in this analysis are not really under control.
A recent work of Pekurovsky and Kilcup \cite{kilcup:99}, in which
$\bsi$ is even found to be negative, unfortunately supports  this
criticism.
We have to conclude that there are no solid predictions for
$B^{(1/2)}_{5,6}$ from the lattice at present.

\begin{table}[thb]
\caption[]{ Lattice results for $B^{(3/2)}_{7,8} (2~\gev)$ obtained
by various groups. 
\label{tab:317}}
\begin{center}
\begin{tabular}{|c|c|c|c|}\hline
  { Fermion type}& $B^{(3/2)}_7$& $B^{(3/2)}_8$ & Matching \\
 \hline
Staggered\cite{GKS}& $0.62(3)(6)$ &$0.77(4)(4)$ & 1-loop \\
Wilson\cite{G67}& $0.58(2)(7)$ &$0.81(3)(3)$ & 1-loop \\
Clover\cite{APE}& $0.58(2)$ &$0.83(2)$ & 1-loop \\
Clover\cite{APE}& $0.72(5)$ &$1.03(3)$ &  Non-pert. \\
\hline
\end{tabular}
\end{center}
\end{table}

\subsubsection{$B^{(1/2)}_{6}$ and $B^{(3/2)}_{8}$ from the 1/N Approach}
The 1/N approach to weak hadronic matrix elements was introduced
in \cite{bardeen:87}. 
In this approach the 1/N expansion becomes a loop expansion
in an effective meson theory. In the strict large-N limit only
the tree level matrix elements of $Q_6$ and $Q_8$ contribute
and one finds (\ref{LN})
while $B^{(1/2)}_{5}=B^{(3/2)}_{7}=0$. The latter fact is not disturbing,
however, as the operators $Q_5$ and $Q_7$  having small Wilson
coefficients are unimportant for $\epe$.

In view of the fact that for $B^{(1/2)}_{6}=B^{(3/2)}_{8}=1$ and the
known value of $\mt$ there is a strong cancellation between
gluon and electroweak penguin contributions to $\epe$, it is
important to investigate whether the $1/N$ corrections significantly
affect this cancellation. 
This has been investigated in \cite{DORT98}, where
a calculation
of $\langle Q_6\rangle_0$ and $\langle Q_8\rangle_2$ in the
twofold expansion in powers of external momenta $p$, and in 
$1/N$ has been presented. The final results for 
$\langle Q_6\rangle_0$ and $\langle Q_8\rangle_2$ 
in \cite{DORT98} include the orders $p^2$ and $p^0/N$.
For $\langle Q_8\rangle_2$ also the term $p^0$ contributes.
Of particular interest are the $\ord(p^0/N)$ contributions
resulting from non-factorizable chiral loops which are
important for the matching between long- and short-distance
contributions. The cut-off scale $\Lambda_c$ in these
non-factorizable diagrams is identified with the QCD renormalization
scale $\mu$ which enters the Wilson coefficients.

\begin{table}[thb]
\caption[]{ Results for $\bsi$ and $\bei$ obtained
in the $1/N$ approach. 
\label{tab:318}}
\begin{center}
\begin{tabular}{|c||c|c|c|c|}\hline
  & $\Lambda_c=0.6~\gev$ & $\Lambda_c=0.7~\gev$ & 
$\Lambda_c=0.8~\gev$ & $\Lambda_c=0.9~\gev$ \\
 \hline\hline
$B_{6}^{(1/2)}$ & $1.10$ &$0.96$ & $0.84$ & $ 0.72 $ \\
                & $(1.30)$ &$(1.19)$ & $(1.09)$ & $(0.99) $ \\
$B_{8}^{(3/2)}$ & $0.64$ &$0.56$ & $0.49$ & $ 0.42 $ \\
                & $(0.71)$ &$(0.65)$ & $(0.59)$ & $(0.53) $ \\
\hline
\end{tabular}
\end{center}
\end{table}

In table \ref{tab:318}, taken from \cite{DORT98,DORT99}, 
we show the values of
$B_{6}^{(1/2)}$ and $B_{8}^{(3/2)}$ as functions of the cut-off
scale $\Lambda_c$. The results depend on whether $F_\pi$ or
$F_K$ is used in the calculation, the difference being of higher
order. The results using $F_K$ are shown  in parentheses.
The decrease of both B-factors with $\Lambda_c=\mu$ is qualitatively
consistent with their $\mu$-dependence found for $\mu\ge 1$ in
\cite{BJLW}, but it is much stronger. Clearly one could also
expect a stronger $\mu$-dependence in the analysis of \cite{BJLW}
for $\mu\le 1~\gev$, but in view of large perturbative corrections
for such small scales a meaningful test of the dependence in
table \ref{tab:318} cannot be made.
We note that for $\Lambda_c=0.7~\gev$  the value of $B_{6}^{(1/2)}$
is close to unity as in the large-N limit. However, $B_{8}^{(3/2)}$
is considerably suppressed. An interesting feature of these
results is the near $\Lambda_c$ independence of the ratio $\bsi/\bei$.
Consequently the results in \cite{DORT98,DORT99} can be summarized by

\be
\label{DOR}
\frac{\bsi}{\bei} \approx 1.72~(1.84),
\qquad 0.72~(0.99)\le\bsi\le 1.10~(1.30)~.
\ee
It is difficult to decide which value should be used in the phenomenology
of $\epe$. On the one hand, for $\Lambda_c\ge 0.7~\gev$ neglected 
contributions from vector mesons in the loops should be included.
On the other hand for $\Lambda_c=\mu= 0.6~\gev$ 
the short distance calculations are questionable.
Probably the best thing to do at present is to vary $\Lambda_c=\mu$
in the full range shown in table \ref{tab:318}.
This has been done in a recent analysis \cite{DORT99b} in
which $\epe$ has been found to be
a decreasing function of $\Lambda_c$.
 
Finally, we would like to mention that the first non-trivial
$1/N$ corrections to the matrix elements of $Q_7$ have been calculated
in \cite{DER1} using the methods developed in \cite{DER2}. In
particular it has been found that $B^{(3/2)}_7$ is a rather
strongly increasing function of $\mu$ with negative values for
$\mu\le\mc$, $B^{(3/2)}_7(\mc)=0$ and positive values for 
$\mu>\mc$. This strong $\mu$-dependence of $B^{(3/2)}_7$ is rather
surprising as the numerical renormalization group analysis
in \cite{BJLW} has shown a rather weak dependence of this 
parameter. We suspect that the inclusion of the full mixing
between $Q_7$ and other operators in the analysis of \cite{DER1}
would weaken the $\mu$-dependence of $B^{(3/2)}_7$
considerably. While this issue requires an additional
investigation, the value of $B^{(3/2)}_7$ has
fortunately only a minor impact on $\epe$. Setting
$B^{(3/2)}_7(\mc)=0$ instead of $B^{(3/2)}_7(\mc)=\bei(\mc)$
used here would change our results for $\epe$ only
by a few percent.

\subsubsection{$B^{(1/2)}_{6}$ and $B^{(3/2)}_{8}$ from 
the Chiral Quark Model}
Effective Quark Models of QCD can be derived in the framework of 
the extended Nambu-Jona-Lasinio model of chiral symmetry 
breaking \cite{NJL}.
For kaon decays and in particular for $\epe$, an extensive analysis
of this model including chiral loops, gluon and $\ord(p^4)$ corrections
has been performed over the last years by the Trieste group
\cite{TR96,TR97}. The
crucial parameters in this approach are a mass parameter $M$ and
the condensates $\langle\bar q q\rangle $ and $\langle\as GG\rangle$.
They can be constrained by imposing the $\Delta I=1/2$ rule.

Since there exists a  nice review \cite{BERT98} by the Trieste 
group, we will only quote here their estimates of the relevant 
$B_i$ parameters. They are given in the HV scheme as follows
\be\label{TRIESTE}
B^{(1/2)}_{6}=1.6\pm0.3, \quad\quad B^{(3/2)}_{8}=0.92\pm 0.02~,
\quad\quad{(\rm Chiral~QM)}.
\ee
Translating these values into the NDR scheme by means of (\ref{NDRHV})
one finds
\be\label{TRHV}
B^{(1/2)}_{6}=1.33\pm0.25, \quad\quad B^{(3/2)}_{8}=0.77\pm 0.02~,
\quad\quad{(\rm NDR)}.
\ee
We observe a substantial enhancement of $B^{(1/2)}_{6}$ in the
chiral quark model, not found in other calculations,
and a
moderate suppression of $B^{(3/2)}_{8}$. The errors given above arise
from the variation of $\ms$. We will return to this point in
subsection \ref{SL}.

It should be remarked that the definitions of the $B_i$ parameters
used in \cite{BERT98} agree with our definitions only if
in the vacuum insertion formulae in \cite{BERT98} the 
$\langle\bar q q\rangle$
condensate is given in terms of $\ms$ as follows:
\be
\langle\bar q q\rangle^2 =\frac{F_\pi^4}{4} 
\left[\frac{m_{\rm K}^2}{\ms + \md}\right]^2.
\ee
This means that in the usual PCAC relation one has to set
$F_K=F_\pi$.

It is interesting to observe that  in this method
$\bsi/\bei=1.74\pm0.33$ in the ball park of (\ref{DOR}).
It will be of interest to see whether future lattice calculations
will confirm this
correlation between $\bsi$ and $\bei$.
 
\subsubsection{$\bsi$ and the $\Delta I=1/2$ Rule}
In  one of the first estimates of $\epe$, Gilman and Wise \cite{GW79}
used the suggestion of Vainshtein, Zakharov and Shifman \cite{PENGUIN}
that the amplitude $\RE A_0$ is dominated by the QCD-penguin operator
$Q_6$. Estimating $\langle Q_6 \rangle_0$ in this manner they predicted
a large value of $\epe$. Since then it has been understood 
\cite{DI12,kilcup:99,DORT99} that
as long as the scale $\mu$ is not much lower than $1~\gev$
the amplitude $\RE A_0$ is dominated by the operators $Q_1$ and
$Q_2$, rather than by $Q_6$. Indeed, at least in the HV
scheme the operator $Q_6$ does not contribute to $\RE A_0$
for $\mu=\mc$ at all, as its coefficient $z_6(\mc)$ relevant for
this amplitude vanishes. Also in the NDR scheme $z_6(\mc)$ is
negligible.

For decreasing $\mu$ the coefficient $z_6(\mu)$ increases and
the $Q_6$ contribution to $\RE A_0$ is larger.
However, if the analyses in \cite{DI12,kilcup:99,DORT99} are
taken into account, the operators $Q_1$ and $Q_2$ are responsible
for at least $90\%$ of $\RE A_0$ if the scale $\mu=1~\gev$ is considered.
Therefore in our opinion there is no strict relation between the large
value of $\epe$ and the $\Delta I=1/2$ rule as sometimes
stated in the literature. 
Moreover, if the $90\%$ contribution of the operators 
$Q_1$ and $Q_2$ to $\RE A_0$ is taken into account and $z_6(1~\gev)$
is calculated in the NDR scheme, $\bsi$ cannot exceed 1.5 if
$\ms(1~\gev)=150\mev$. Consequently we do not think that values of
$\bsi$ in the NDR scheme as high as 4.0 suggested in \cite{Nierste}
are plausible. Unfortunately, due to the very strong $\mu$ and
renormalization scheme dependences of $z_6(\mu)$, 
general
definite conclusions about $\bsi$ cannot be reached in this manner
at present.
Similarly,
we cannot exclude the possibility that $\bsi$ is substantially
higher than unity if it turned out that the present methods
overestimate the role of $Q_1$ and $Q_2$ in $\RE A_0$.

\subsubsection{Summary}
We have seen that most non-perturbative approaches discussed
above found $\bei$ below unity. The suppression of $\bei$
below unity is rather modest (at most $20\%$) in the lattice 
approaches and in the chiral quark model. In the $1/N$ approach
$\bei$ is rather strongly suppressed and can be as low as 0.5.

Concerning $\bsi$ the situation is worse. As we stated above
there is no solid prediction for this parameter in the lattice
approach. On the other hand while the average value of $\bsi$
in the $1/N$ approach is close to $1.0$, the chiral
quark model gives at $\mu=0.8~\gev$ and in the NDR scheme
the value for $\bsi$
as high as $1.33\pm 0.25$.
Interestingly both approaches give the ratio
$\bsi/\bei$ in the ball park of 1.7.

Guided by the results presented above and biased to some
extent by the results from the large-N approach and lattice
calculations, we will use
in our numerical analysis below $\bsi$ and $\bei$ in
the ranges:
\be\label{bbb}
\bsi=1.0\pm0.3,
\qquad
\bei=0.8\pm 0.2
\ee
keeping always $\bsi\ge \bei$.
In our 1996 analysis \cite{BJL96a} we have used $\bsi=1.0\pm0.2$ and 
$\bei=1.0\pm0.2$ without the constraint $\bsi\ge \bei$. 
The decrease of $\bei$ below unity is motivated 
by the recent results discussed above. The increase in the range
of $\bsi$ is supposed to take effectively into account the
uncertainty in $\OEE$ which we estimate to be at most $\pm 30\%$
i.e $\OEE=0.25\pm0.08$. We will return to this point in Section 3.
\section{Numerical Results in the Standard Model}\label{sec:standard}
\subsection{Input Parameters}
In order to make predictions for $\epe$ we need the value of
$\IM \lambda_t$. This can be obtained from the standard analysis
of the unitarity triangle which uses the data for $\vcb$, $|V_{ub}|$,
$\eps$, $\Delta M_d$ and $\Delta M_s$, where the last two
measure the size of
$B^0_{d,s}-\bar B^0_{d,s}$ mixings.
Since this analysis is very well known we do not list the
relevant formulae here. They can be found for instance 
in \cite{AJBLH,UT99}.

The input parameters needed to perform the
standard analysis of the unitarity triangle
are given in table \ref{tab:inputparams}, where
 $\mt$ 
 refers
to the running current top quark mass defined at $\mu=\mt^{Pole}$.
It corresponds to 
$\mt^{Pole}=174.3\pm 5.1\gev$ measured by CDF and D0 \cite{CDFD0}.

We also recall that
the lower bound on $\Delta M_s$
together with $\Delta M_d$  puts the following constraint on
the ratio $\vtd/|V_{ts}|$:
\begin{equation}\label{107b}
\frac{\vtd}{|V_{ts}|}< 
\xi\sqrt{\frac{m_{B_s}}{m_{B_d}}}
\sqrt{\frac{\Delta M_d}{\Delta M^{\rm min}_s}},
\qquad
\xi = 
\frac{F_{B_s} \sqrt{B_{B_s}}}{F_{B_d} \sqrt{B_{B_d}}}.
\end{equation}

The range for $\Lms^{(4)}$ in table~\ref{tab:inputparams} 
corresponds roughly to $\alpha_s(\mz)=0.119\pm 0.003$.

\begin{table}[thb]
\caption[]{Collection of input parameters.
We impose $\bsi\ge\bei$.
\label{tab:inputparams}}
\vspace{0.4cm}
\begin{center}
\begin{tabular}{|c|c|c|c|}
\hline
{\bf Quantity} & {\bf Central} & {\bf Error} & {\bf Reference} \\
\hline
$|V_{cb}|$ & 0.040 & $\pm 0.002$ & \cite{PDG}     \\
$|V_{ub}|$ & $3.56\cdot 10^{-3}$ & $\pm 0.56\cdot 10^{-3} $ &
\cite{STOCCHI}  \\
$\hat B_K$ & 0.80 & $\pm 0.15$ & See Text  \\
$\sqrt{B_d} F_{B_{d}}$ & $200\mev$ & $\pm 40\mev$ & \cite{BF} \\
$\mt$ & $165\gev$ & $\pm 5\gev$ & \cite{CDFD0} \\
$\Delta M_d$ & $0.471~\mbox{ps}^{-1}$ & $\pm 0.016~\mbox{ps}^{-1}$ 
& \cite{LEPB}\\
$\Delta M_s$ & $>12.4~\mbox{ps}^{-1}$ & $ 95\% {\rm C.L.}$ 
& \cite{LEPB}\\
$\xi$ & $1.14$ & $\pm 0.08$ 
& \cite{BF} \\
$\Lms^{(4)}$ & $340 \mev$ & $\pm 50\mev$ & \cite{PDG,BETKE} \\
$\ms(\mc)$ & $130\mev$    & $\pm 25\mev$ & See Text\\
$\bsi $ & 1.0 & $\pm 0.3$ & See Text\\
$\bei $ & 0.8 & $\pm 0.2$ & See Text\\
\hline
\end{tabular}
\end{center}
\end{table}

\subsection{Monte Carlo and Scanning Estimates of $\epe$}

In what follows we will present two types of  numerical analyses of
${\rm Im}\lambda_t$ and $\epe$:

\begin{itemize}
\item
Method 1: The experimentally measured numbers  are used with Gaussian errors
 and for the theoretical input parameters we take a flat distribution
in the ranges given in
table~\ref{tab:inputparams}.
\item
Method 2: Both the experimentally measured numbers and the theoretical input
parameters are scanned independently within the ranges given in
table~\ref{tab:inputparams}.
\end{itemize}

\begin{figure}
\begin{center}
% GNUPLOT: LaTeX picture with Postscript
\setlength{\unitlength}{0.1bp}
\begin{picture}(4539,2808)(0,0)
\special{psfile=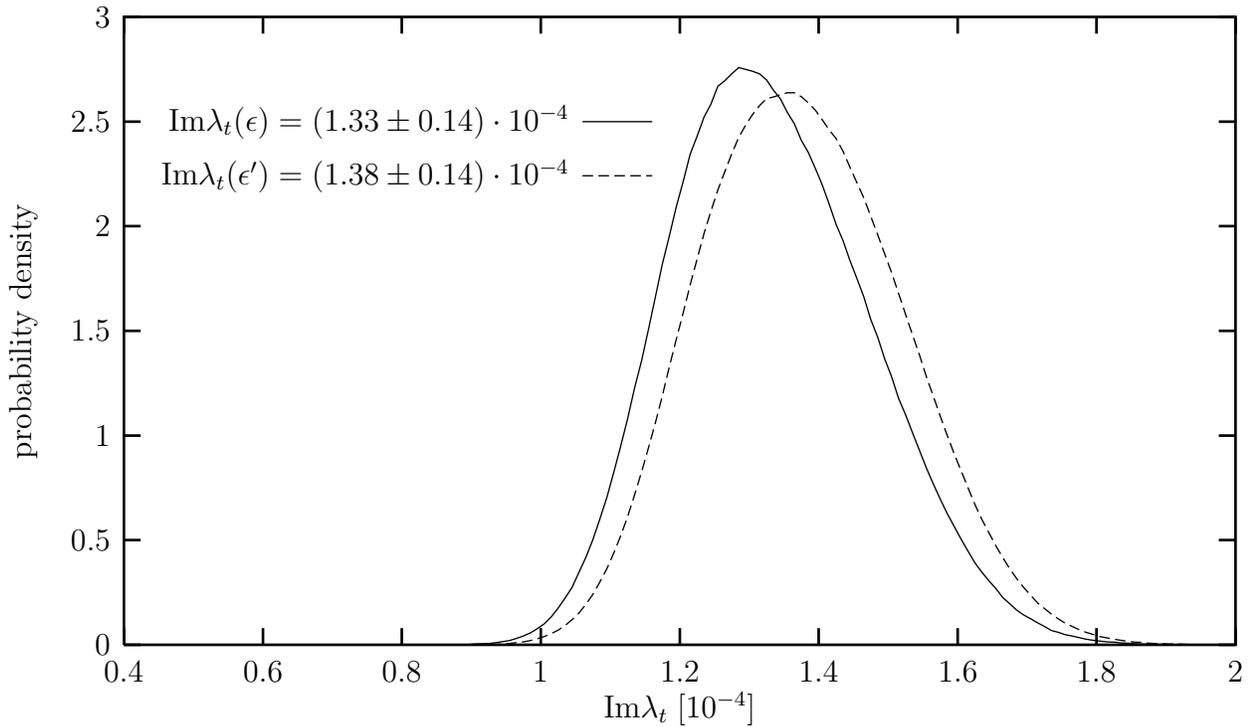 llx=0 lly=0 urx=908 ury=655 rwi=9080}
\put(2028,2109){\makebox(0,0)[r]{Im$\lambda_t(\epsilon')= (1.38\pm0.14)\cdot 10^{-4}$}}
\put(2028,2313){\makebox(0,0)[r]{Im$\lambda_t(\epsilon)= (1.33\pm0.14)\cdot 10^{-4}$}}
\put(2444,100){\makebox(0,0){Im$\lambda_t\;[10^{-4}]$}}
\put(0,1524){%
\special{ps: gsave currentpoint currentpoint translate
270 rotate neg exch neg exch translate}%
\makebox(0,0)[b]{\shortstack{probability density}}%
\special{ps: currentpoint grestore moveto}%
}
\put(4539,240){\makebox(0,0){2}}
\put(4015,240){\makebox(0,0){1.8}}
\put(3492,240){\makebox(0,0){1.6}}
\put(2968,240){\makebox(0,0){1.4}}
\put(2445,240){\makebox(0,0){1.2}}
\put(1921,240){\makebox(0,0){1}}
\put(1397,240){\makebox(0,0){0.8}}
\put(874,240){\makebox(0,0){0.6}}
\put(350,240){\makebox(0,0){0.4}}
\put(300,2708){\makebox(0,0)[r]{3}}
\put(300,2313){\makebox(0,0)[r]{2.5}}
\put(300,1919){\makebox(0,0)[r]{2}}
\put(300,1524){\makebox(0,0)[r]{1.5}}
\put(300,1129){\makebox(0,0)[r]{1}}
\put(300,735){\makebox(0,0)[r]{0.5}}
\put(300,340){\makebox(0,0)[r]{0}}
\end{picture}
\end{center}
\vspace{-6mm}
\caption{Probability density distributions for $\IM\lambda_t$ without
(solid line)
and with (dashed line) the $\epe$-constraint.}
\label{g3}
\end{figure}

\begin{figure}
\begin{center}
% GNUPLOT: LaTeX picture with Postscript
\setlength{\unitlength}{0.1bp}
\begin{picture}(4539,2808)(0,0)
\special{psfile=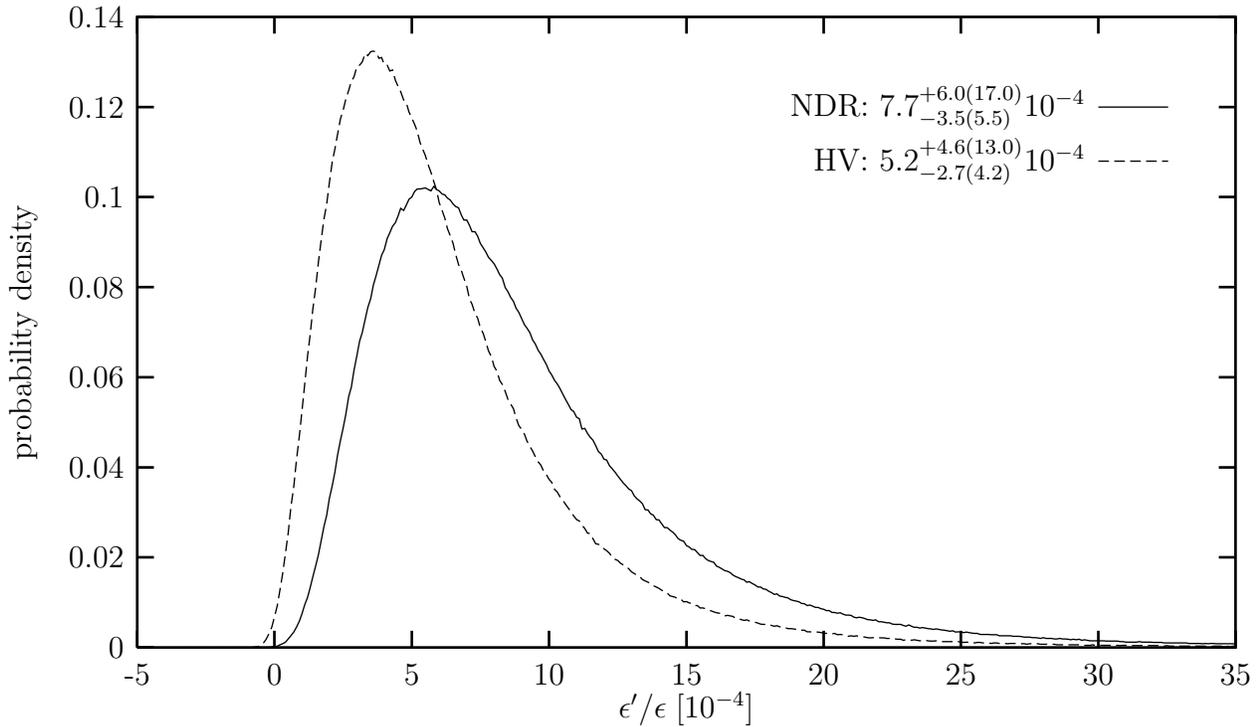 llx=0 lly=0 urx=908 ury=655 rwi=9080}
\put(3972,2164){\makebox(0,0)[r]{HV: $5.2^{+4.6(13.0)}_{-2.7(4.2)} 10^{-4}$}}
\put(3972,2368){\makebox(0,0)[r]{NDR: $7.7^{+6.0(17.0)}_{-3.5(5.5)} 10^{-4}$}}
\put(2469,100){\makebox(0,0){$\epsilon'/\epsilon \;[10^{-4}]$}}
\put(0,1519){%
\special{ps: gsave currentpoint currentpoint translate
270 rotate neg exch neg exch translate}%
\makebox(0,0)[b]{\shortstack{probability density}}%
\special{ps: currentpoint grestore moveto}%
}
\put(4539,230){\makebox(0,0){35}}
\put(4022,230){\makebox(0,0){30}}
\put(3504,230){\makebox(0,0){25}}
\put(2987,230){\makebox(0,0){20}}
\put(2470,230){\makebox(0,0){15}}
\put(1952,230){\makebox(0,0){10}}
\put(1435,230){\makebox(0,0){5}}
\put(917,230){\makebox(0,0){0}}
\put(400,230){\makebox(0,0){-5}}
\put(350,2708){\makebox(0,0)[r]{0.14}}
\put(350,2368){\makebox(0,0)[r]{0.12}}
\put(350,2029){\makebox(0,0)[r]{0.1}}
\put(350,1689){\makebox(0,0)[r]{0.08}}
\put(350,1349){\makebox(0,0)[r]{0.06}}
\put(350,1009){\makebox(0,0)[r]{0.04}}
\put(350,670){\makebox(0,0)[r]{0.02}}
\put(350,330){\makebox(0,0)[r]{0}}
\end{picture}
\end{center}
\vspace{-6mm}
\caption{Probability density distributions for $\epe$ in NDR and HV
schemes.}
\label{g1}
\end{figure}

Using the first method we find the probability density 
distributions for
${\rm Im}\lambda_t$ and $\epe$ in figs.~\ref{g3} and \ref{g1}
 respectively. 
From the distributions in figs.~\ref{g3} and \ref{g1}
we deduce the following results:
\begin{equation}
 {\rm Im}\lambda_t =( 1.33\pm 0.14) \cdot 10^{-4}
\label{eq:imfinal}
\end{equation}
\begin{equation}
~~~~~~\epe= ( 7.7^{~+6.0}_{~-3.5}) \cdot 10^{-4}\qquad {\rm (NDR)}
\label{eq:eperangefinal}
\end{equation}

Since the probability density in fig.~\ref{g3} is rather 
symmetric we give only the
mean and the standard deviation for $\IM\lambda_t$.
On the other hand, the resulting probability density distribution for
$\epsilon'/\epsilon$ is very asymmetric with a very long
tail towards large values. Therefore we decided  to quote the median
and the $68\%(95\%)$ confidence level intervals. This means that $68\%$
of our data can be found inside the corresponding error interval and
that $50\%$ of our data has smaller $\epsilon'/\epsilon$ than our
median.

We observe that negative values of $\epsilon'/\epsilon$ can be
excluded at $95\%$ C.L. For completeness we quote the mean and the
standard deviation for $\epsilon'/\epsilon$:
\begin{equation}\label{mean}
\epe=9.1\pm6.2 \qquad {\rm (NDR)}
\end{equation}

Using the second method and the parameters in table~\ref{tab:inputparams} 
we find :
\begin{equation}
1.04 \cdot 10^{-4} \le {\rm Im}\lambda_t \le 1.63 \cdot 10^{-4}
\label{eq:imnew}
\end{equation}
\begin{equation}
~~~~~1.05 \cdot 10^{-4} \le \epe \le 28.8 \cdot 10^{-4}\qquad {\rm (NDR)}.
\label{eq:eperangenew}
\end{equation}

The above results for $\epe$ apply to the NDR scheme.
$\epe$ is generally lower in the HV scheme if the same values for
$B_6^{(1/2)}$ and $B_8^{(3/2)}$ are used in both schemes. 
As discussed in subsection \ref{rsd}, such treatment of
$B_6^{(1/2)}$ and $B_8^{(3/2)}$ is the proper way of estimating
scheme dependences at present.
 
Using the two error analyses we find respectively:
\begin{equation}
~~~~~~\epe= ( 5.2^{~+4.6}_{~-2.7}) \cdot 10^{-4}\qquad {\rm (HV)}
\label{hv:final}
\end{equation}
and
\begin{equation}
~~~~~0.26 \cdot 10^{-4} \le \epe \le 22.0 \cdot 10^{-4}\qquad {\rm (HV)}.
\label{hv:eperangenew}
\end{equation}
Moreover, the mean and the standard deviation read
\begin{equation}\label{hmean}
\epe  =6.3\pm4.8 \qquad {\rm (HV)}.
\end{equation}
The corresponding probability density distribution for $\epe$ is compared 
to the one obtained in the NDR scheme in  fig.~\ref{g1}.
Assuming, on the other hand, that the values in (\ref{bbb}) correspond
to the NDR scheme and using the relation (\ref{NDRHV}), we find for
the HV scheme the range 
$0.58 \cdot 10^{-4} \le \epe \le 26.9 \cdot 10^{-4}$ which is
much closer to the NDR result in (\ref{eq:eperangenew}). 
This exercise shows that it is very desirable to have the
scheme dependence under control.

We observe that the most probable values for $\epe$ in the NDR scheme
 are in the
ball park of $1 \cdot 10^{-3}$. They are lower by roughly $30\%$ in the
HV scheme if the same values for $(\bsi,\bei)$ are used.
On the other hand the ranges in (\ref{eq:eperangenew}) and
(\ref{hv:eperangenew}) show that for
particular choices of the input parameters, values for $\epe$ as high as
$(2-3)\cdot 10^{-3}$ cannot be excluded at present. Let us study 
this in  more detail.

\subsection{Anatomy of $\epe$}
\subsubsection{Global Analysis}
In table~ \ref{tab:31731} we show the values 
 of $\epe$ in units of $10^{-4}$ 
for specific values of $\bsi$, $\bei$ and $\ms(\mc)$ as calculated
in the NDR scheme. The corresponding values in the HV scheme
are lower as discussed above.
The fourth column shows the results for central values of all remaining
parameters. The comparison of the the fourth and the fifth column
 demonstrates
how $\epe$ is increased when $\Lms^{(4)}$ is raised from $340~\mev$
to $390~\mev$. As stated in (\ref{ap}) $\epe$ is roughly proportional
to $\Lms^{(4)}$. Finally, in the last column maximal values of $\epe$
are given.
To this end we have scanned all parameters relevant for
the analysis of $\IM\lambda_t$ within one
standard deviation and have chosen the highest value of
$\Lms^{(4)}=390\mev$. Comparison of the last two columns demonstrates
the impact of the increase of $\IM\lambda_t$ from its
central to its maximal value and of the variation of $\mt$.

Table~\ref{tab:31731} gives a good insight in the dependence of
$\epe$ on various parameters which is roughly described
by (\ref{ap}). We observe the following hierarchies:

\begin{itemize}
\item
The largest uncertainties reside in $\ms$, $\bsi$ and $\bei$.
$\epe$ increases universally by roughly a factor of 2.3 when
$\ms(\mc)$ is changed from $155 \mev$ to $105 \mev$. The increase
of $\bsi$ from 1.0 to 1.3 increases $\epe$ by $(55\pm 10)\%$,
depending on $\ms$ and $\bei$. The corresponding changes
due to $\bei$ are approximately $(40\pm 15)\%$.
\item
The combined uncertainty due to $\IM\lambda_t$ and $\mt$,
present both in $\IM\lambda_t$ and $F_{\varepsilon'}$,
is approximately $\pm 25\%$. The uncertainty due to
$\mt$ alone is only $\pm 5\%$.
\item
The uncertainty due to $\Lms^{(4)}$ is approximately $\pm 16\%$.
\end{itemize}

The large sensitivity of $\epe$ to $\ms$ has been known 
since the analyses in the eighties. In the context of
the KTeV result this issue has been analyzed in \cite{Nierste}.
It has been found that provided $2\bsi-\bei\le 2$ the
consistency of the Standard Model with the KTeV result
requires the $2\sigma$ bound $\ms(2\gev)\le 110\mev$.
Our analysis is compatible with these findings.

It is of interest to investigate the impact of the 
relation (\ref{DOR}) on our results. Scanning
all parameters in the ranges given in table~\ref{tab:inputparams} 
and imposing $\bsi=1.7\cdot \bei$ we find
\begin{equation}
3.7 \cdot 10^{-4} \le \epe \le 26.2 \cdot 10^{-4}
\label{eperangenew}
\ee
which is somewhat reduced with respect to (\ref{eq:eperangenew}).

Finally we would like to comment on formula (\ref{epeth1})
in which $\RE\lambda_t$ appears instead of $\IM\lambda_t$.
Since $F_{\varepsilon}$  decreases with decreasing $\RE\lambda_t$
one can come closer to the experimental data for $\epe$ by choosing
$\RE\lambda_t$ sufficiently small. In the 
Wolfenstein parametrization $\RE\lambda_t$ is proportional
to $1-\varrho$  and a small $\RE\lambda_t$  corresponds to a
 sufficiently
large positive value of the parameter $\varrho$. Yet it is known
from analyses of the unitarity triangle that $\varrho$ is bounded
from above by the ratio $\vub$ and even stronger by the value of
$\eps$. If these constraints are taken into account the analysis
using (\ref{epeth1}) reduces to the one presented above.

\begin{table}[thb]
\caption[]{ Values of $\epe$ in units of $10^{-4}$ 
for specific values of $\bsi$, $\bei$ and $\ms(\mc)$
and other parameters
as explained in the text.
\label{tab:31731}}
\begin{center}
\begin{tabular}{|c|c|c|c|c||c|}\hline
 $B^{(1/2)}_6$& $B^{(3/2)}_8$ & $\ms(\mc)[\mev]$ &
  Central & $\Lms^{(4)}=390\mev $ & Maximal \\ \hline
      &     & $105$ &  20.2 & 23.3 & 28.8\\
 $1.3$&$0.6$& $130$ &  12.8 & 14.8 & 18.3\\
      &     & $155$ &   8.5 &  9.9 & 12.3 \\
 \hline
      &     & $105$ &  18.1 & 20.8 & 26.0\\
 $1.3$&$0.8$ & $130$ & 11.3 & 13.1 & 16.4\\
      &     & $155$ &   7.5 &  8.7 & 10.9\\
 \hline
      &     & $105$ &   15.9 & 18.3 & 23.2\\
 $1.3$&$1.0$ & $130$ &  9.9 &  11.5 & 14.5\\
      &     & $155$ &   6.5  &  7.6 &  9.6\\
 \hline\hline
      &     & $105$ &   13.7 & 15.8 & 19.7\\
 $1.0$&$0.6$ & $130$ &  8.4 &  9.8& 12.2 \\
      &     & $155$ &   5.4 &  6.4 & 7.9 \\
 \hline
      &     & $105$ &   11.5 & 13.3 & 16.9\\
$1.0$&$0.8$ & $130$  &  7.0 &   8.1 & 10.4\\
     &     & $155$ &   4.4 &    5.2 &  6.6\\
 \hline
     &     & $105$ &   9.4 &   10.9 & 14.1 \\
$1.0$&$1.0$ & $130$  &  5.5 &   6.5 &  8.5 \\
     &     & $155$ &   3.3 &    4.0 &  5.2\\
 \hline
\end{tabular}
\end{center}
\end{table}

\subsubsection{Parametric vs. Hadronic Uncertainties}
One should distinguish between parametric and hadronic uncertainties.
Parametric uncertainties are related to $\mt$, $|V_{ub}|$, $\vcb$
and $\Lms^{(4)}$. One should in principle include $\ms$ in this list.
However, in order to extract $\ms$ from the kaon mass one
encounters large non-perturbative uncertainties. Clearly such
uncertainties are also present in the determination of $\vcb$ and
in particular in the determination of $|V_{ub}|$, 
but they are substantially smaller.
Hence the hadronic uncertainties discussed below are related to
$\hat B_K$, $\bsi$, $\bei$ and $\ms$.

In table~\ref{tab:3} 
we show ranges for $\epe$ related to various uncertainties.
The parametric uncertainties have been obtained for central values of
$\hat B_K$ and $\ms$ and two choices of $(\bsi,\bei)$. The hadronic 
uncertainties due to $\hat B_K$, $\bsi$ and $\bei$ have been found
by setting all the remaining parameters at their central values.
The uncertainty due to $\ms$ has been shown for
two choices of $(\bsi,\bei)$ and all other parameters set at 
their central
values. The last row in table~\ref{tab:3} 
shows the total hadronic uncertainty. It is evident from this 
table that hadronic uncertainties dominate, although the
reduction of parametric uncertainties is very desirable.

\begin{table}[thb]
\caption[]{ Uncertainties in $\epe$ in units of $10^{-4}$ 
as explained in the text.
\label{tab:3}}
\begin{center}
\begin{tabular}{|c|c|c|c|c|}\hline
 Uncertainties & $\bsi$ & $B^{(3/2)}_8$ & $(\epe)_{\rm min}$&
 $(\epe)_{\rm max}$ \\ \hline
 Parametric & 1.0 &  0.8 &  5.0 & 9.5 \\
 Parametric & 1.3 &  0.8 &  8.4 & 15.1 \\
\hline 
Hadronic ($B_i$) &  --   & -- & 3.0 & 13.6 \\
Hadronic ($\ms$) & 1.0  & 0.8 & 4.5 & 11.3 \\
Hadronic ($\ms$) & 1.3  & 0.8 & 7.6 & 17.9 \\
Hadronic (full ) &  --   & -- & 1.7 & 21.3 \\
 \hline
\end{tabular}
\end{center}
\end{table}

\subsection{$\bsi$-$\bei$ Plot}
In fig.~\ref{g2} we show the minimal value of $\bsi$ for two
choices of $\ms(\mc)$ and $\Lms^{(4)}$
as a function of $\bei$ for which the theoretical value of
$\epe$ is higher than $2.0 \cdot 10^{-3}$. To obtain
this plot we have varied all other parameters in
the ranges given in table~\ref{tab:inputparams}.
We show also the
line corresponding to the relation (\ref{DOR}).
We observe that as long as $\bei\ge0.6$, the parameter
$\bsi$ is required to be larger than unity.
This plot should be useful when our knowledge of
$\bsi$, $\bei$, $\ms$ and $\Lms^{(4)}$ improves.

\begin{figure}
\begin{center}
% GNUPLOT: LaTeX picture with Postscript
\setlength{\unitlength}{0.1bp}
\begin{picture}(4539,2808)(0,0)
\special{psfile=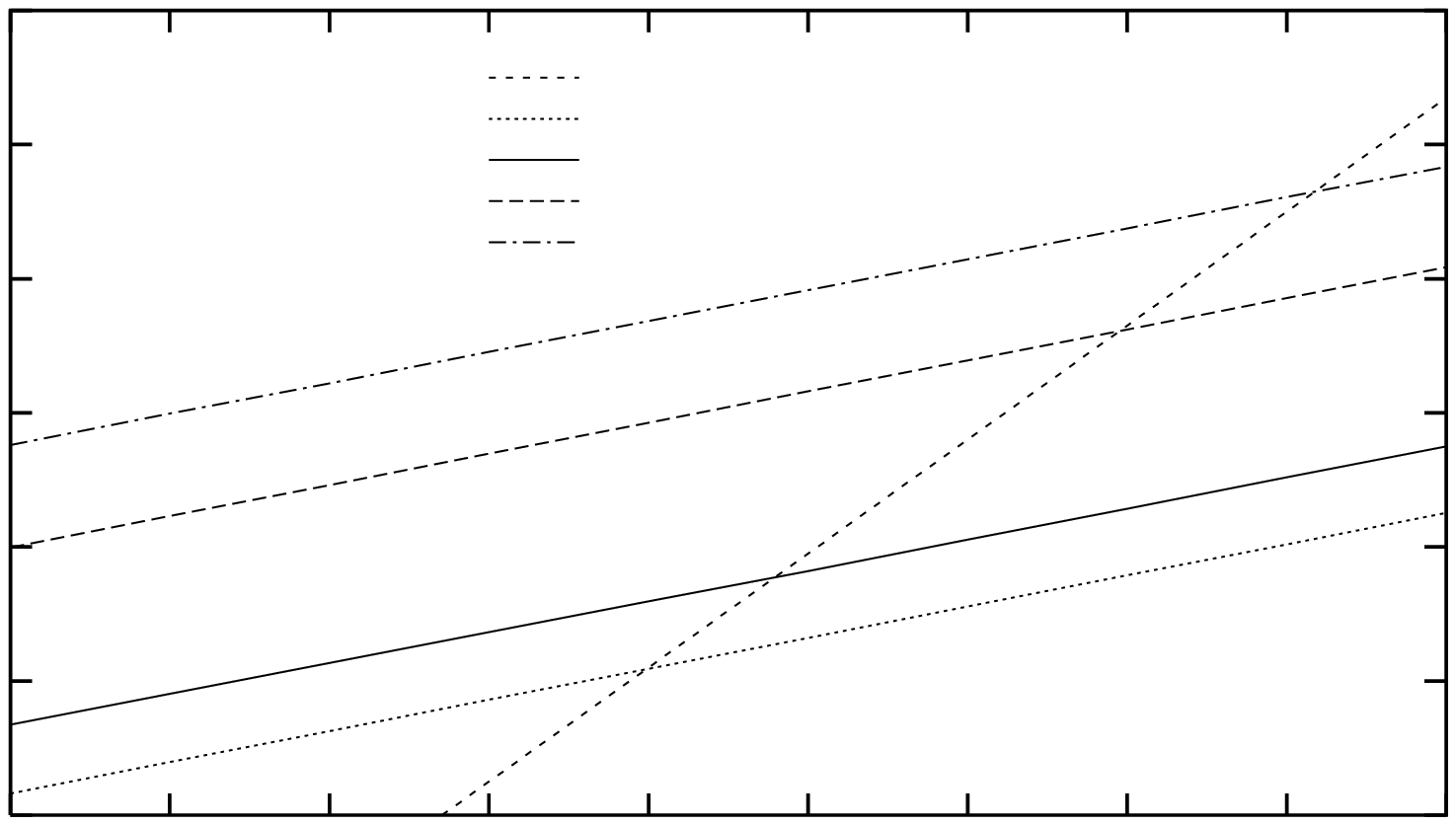 llx=0 lly=0 urx=908 ury=655 rwi=9080}
\put(1696,2032){\makebox(0,0)[r]{$\scriptstyle m_s=130{\rm MeV} \quad \Lambda_{\overline{\rm MS}} \, =340{\rm MeV}$}}
\put(1696,2152){\makebox(0,0)[r]{$\scriptstyle m_s=130{\rm MeV} \quad \Lambda_{\overline{\rm MS}} \, =390{\rm MeV}$}}
\put(1696,2272){\makebox(0,0)[r]{$\scriptstyle m_s=105{\rm MeV} \quad \Lambda_{\overline{\rm MS}} \, =340{\rm MeV}$}}
\put(1696,2392){\makebox(0,0)[r]{$\scriptstyle m_s=105{\rm MeV} \quad \Lambda_{\overline{\rm MS}} \, =390{\rm MeV}$}}
\put(1696,2512){\makebox(0,0)[r]{$\scriptstyle B_6=1.7B_8$}}
\put(2444,100){\makebox(0,0){$B_8^{(3/2)}$}}
\put(0,1534){%
\special{ps: gsave currentpoint currentpoint translate
270 rotate neg exch neg exch translate}%
\makebox(0,0)[b]{\shortstack{$B_6^{(1/2)}$}}%
\special{ps: currentpoint grestore moveto}%
}
\put(4539,260){\makebox(0,0){1.1}}
\put(4074,260){\makebox(0,0){1}}
\put(3608,260){\makebox(0,0){0.9}}
\put(3143,260){\makebox(0,0){0.8}}
\put(2677,260){\makebox(0,0){0.7}}
\put(2212,260){\makebox(0,0){0.6}}
\put(1746,260){\makebox(0,0){0.5}}
\put(1281,260){\makebox(0,0){0.4}}
\put(815,260){\makebox(0,0){0.3}}
\put(350,260){\makebox(0,0){0.2}}
\put(300,2708){\makebox(0,0)[r]{2}}
\put(300,2317){\makebox(0,0)[r]{1.8}}
\put(300,1925){\makebox(0,0)[r]{1.6}}
\put(300,1534){\makebox(0,0)[r]{1.4}}
\put(300,1143){\makebox(0,0)[r]{1.2}}
\put(300,751){\makebox(0,0)[r]{1}}
\put(300,360){\makebox(0,0)[r]{0.8}}
\end{picture}
\end{center}
\vspace{-6mm}
\caption{Minimal value of $\bsi$ consistent 
with $\epe\ge 2.0\cdot 10^{-3}$.}
\label{g2}
\end{figure}

\subsection{Approximate Scaling Laws for $\epe$}\label{SL}
\subsubsection{Preliminaries}
Table~\ref{tab:31731} contains a lot of information on $\epe$.
This information can be further extended by noting that $\epe$
depends to a very good approximation on certain combinations
of the input parameters. 
This is seen in (\ref{ap}) and (\ref{eq:pbePi}). 
Here we want to provide scaling laws based on these formulae
which allow to obtain from table~\ref{tab:31731} values for
$\epe$ for different sets of input parameters.
\subsubsection{$\bsi$, $\bei$ and $\ms$}
As seen in (\ref{eq:pbePi}), $\epe$ depends on these important
three parameters only through $R_6$ and $R_8$ defined in 
(\ref{RS}).
Using this property one can for instance immediately find that the
values for $\epe$ in the  tenth row of 
table~\ref{tab:31731} can also be obtained for the set
\be
\bsi=1.50,\qquad \bei=0.90,\qquad \ms(\mc)=130~\mev~.
\ee
This set of parameters is similar
to the input parameters used by the Trieste group \cite{BERT98}.

At this point we would like to remark that in 
principle the determination of $\bsi$ and $\bei$ in a given
non-perturbative framework could depend on the value of $\ms$.
This turns out not to be the case in the large-N approach
\cite{bardeen:87,DORT98,DORT99}. In the lattice approach this
question has still to be investigated. On the other hand
there are results in the literature showing a strong 
$m_s$-dependence of the $B_i$ parameters. This is the case
for $\bsi$ in the chiral quark model where $\bsi$ scales
like $\ms$ \cite{BERT98}. Similarly values for $B_7^{(3/2)}$
calculated in \cite{DER1} show a strong $\ms$-dependence.

In the present paper we have varied $(\bsi,\bei)$ and $\ms$ 
independently which is in accordance with large-N calculations.
This resulted in the following ranges for $R_6$ and $R_8$
\be
0.5\le R_6\le 1.95, \qquad 0.4\le R_8\le 1.5
\ee
which are correlated through their common dependence on $\ms$.

If $(\bsi,\bei)$ depend on $\ms$ these ranges could
change. 
In this context
one should remark that in the chiral quark model
\cite{BERT98} the highest value of $R_6$
corresponds to the minimal value of $\bsi$ and consequently
the comparison of the results from the chiral
quark model and the large-N approach has to be made with
care.

Finally, it should be remarked that the decomposition of 
the relevant hadronic
matrix elements of penguin operators into a product of $B_i$ factors times
$1/m_s^2$ although useful in the $1/N$ approach will become unnecessary in 
the lattice approach, once matrix elements of dimension three
will be calculable with improved accuracy.

\subsubsection{$\Lms^{(4)}$ and $\IM\lambda_t$}
For $\alpha_s(\mz)=0.119\pm0.003$, the ratio $\epe$ is within
a few percent proportional to $\Lms^{(4)}$. On the
 other hand $\epe$ is exactly
proportional to $\IM\lambda_t$ at fixed $\mt$. 
However, if $\mt$ is varied
the correlation in $\mt$ between $\IM\lambda_t$
extracted from $\eps$ and $F_{\varepsilon'}$ has to be taken
into account. Consequently the simple
rescaling of $\epe$ with the values of $\IM\lambda_t$
is only true within a few percent.
\subsubsection{Sensitivity to $\OEE$}
The dependence of $\epe$ on $\OEE$ can be studied numerically by 
using the formula (\ref{eq:P12}) or incorporated approximately
into the analytic formula (\ref{eq:3b}) by simply replacing
$\bsi$ with an effective parameter
\be\label{eff}
(\bsi)_{\rm eff}=\bsi\frac{(1-0.9~\OEE)}{0.775}
\ee
A numerical analysis shows that using $(1-\OEE)$
overestimates the role of $\OEE$. In our numerical analysis
we have incorporated the uncertainty in $\OEE$ by increasing
the error in $\bsi$ from $\pm 0.2$ to $\pm 0.3$.

The last estimates of $\OEE$ have been done more than ten years
ago \cite{donoghueetal:86}-\cite{lusignoli:89} 
and it is desirable to update these analyses which
 can be summarized by
\be
\OEE=0.25\pm 0.08~.
\ee
The uncertainty in $\epe$ due to
$\OEE$ alone is approximately $\pm 12\%$
and is slightly lower than the one
originating
from $\Lms^{(4)}$.
\subsubsection{Sensitivity to $\hat B_K$}
As $\IM\lambda_t$ extracted from $\eps$ increases with
decreasing $\hat B_K$, 
there is a possibility of increasing $\epe$ by
decreasing $\hat B_K$ below the range considered 
in table~\ref{tab:inputparams}. 
It should be remarked that $\epe$ is not simply proportional
to $1/\hat B_K$ as the extraction of $\IM\lambda_t$ from $\eps$
involves also $\RE\lambda_t$ (see (\ref{epsm})). For the phase 
$\delta$ in the first
quadrant as favoured by the analyses
of the unitarity triangle \cite{UT99}, the dependence
of $\epe$ on $\hat B_K$ is weaker than $1/\hat B_K$ \cite{PBE0}.

Now, the highest value of $\IM\lambda_t$
consistent with the unitarity of the CKM matrix is
$1.73\cdot 10^{-4}$. It is obtained from $\eps$ for $\hat B_K=0.52$.
This increase of $\IM\lambda_t$ beyond the range in (\ref{eq:imnew})
 would increase the maximal values
in table~\ref{tab:31731} by approximately $6\%$. On the other hand it
should be emphasized that for $\hat B_K=0.8-0.9$, as indicated
by lattice calculations, $\epe$ is generally smaller than found
in our paper unless $\bsi$ is substantially increased.
This is what happens in the chiral quark model \cite{BERT98}
where on the one hand $\hat B_K=1.1\pm 0.2$ and on the other
hand $\bsi=1.6\pm0.3$.

\subsection{Impact on $\IM\lambda_t$ and the Unitarity Triangle}
As we stressed at the beginning of this paper the main new
parameter to be fitted by means of $\epe$ is $\IM\lambda_t$.
Our analysis indicates that the Standard Model estimates of $\epe$ are
generally below the data. If the parameters $\ms$, $\bsi$, $\bei$,
$\Lms^{(4)}$ and $\OEE$ are such that $\IM\lambda_t$ consistent
with $\eps$ (see (\ref{eq:imnew})) cannot accomodate the
experimental value of $\epe$, one has to conclude that  new
contributions from new physics are required. On the other hand
if the data on $\epe$ can be reproduced within the Standard Model,
then generally a lower bound on $\IM\lambda_t$ excluding
a large fraction of the range (\ref{eq:imnew}) can be obtained.
Unfortunately, the strong dependence of the lower bound on
the parameters involved precludes any firm conclusions.
Similar comments apply to the possible impact of $\epe$ on
the analysis of the unitarity triangle: the presently allowed
area in the $(\bar\varrho,\bar\eta)$ plane \cite{UT99} 
can be totally removed
or an improved lower limit on $\bar\eta$ from $\epe$ will decrease 
the allowed region
considerably.
As an illustration we show in table~\ref{tab:355}
the lower bound on $\IM\lambda_t$ from
$\epe$  as a function of $\bei$ for 
$\ms(\mc)=105~\mev$, $\bsi=1.3$ and $\Lms^{(4)}=390~\mev$. 
To this end we have used
the formula (\ref{eq:epe1}) with $\epe\ge 2.0\cdot 10^{-3}$.
Comparing with (\ref{eq:imnew}) we indeed observe that the lower
bound on $\IM\lambda_t$ has been improved.

The impact of $\epe$-data as given in (\ref{ga}) on $\IM\lambda_t$ 
can also be investigated by the method 1 which was used
to obtain (\ref{eq:imfinal}) and (\ref{eq:eperangefinal}). We
find
\begin{equation}
 {\rm Im}\lambda_t =( 1.38\pm0.14) \cdot 10^{-4}.
\label{mfinal}
\end{equation}
The corresponding distribution is compared with the one without
$\epe$-constraint in fig.~\ref{g3}. 
We observe a very modest but visible shift
towards higher values for $\IM\lambda_t$.

\begin{table}[thb]
\caption[]{ Minimal values of $\IM\lambda_t$,
$Br(K_L\to\pi^0\nu\bar\nu)$ and 
$Br(K_L\to\pi^0 e^+e^-)_{\rm dir}$
for $\ms(\mc)=105~\mev$, $\bsi=1.3$ and $\Lms^{(4)}=390~\mev$ and
 specific values of  $\bei$ 
assuming $\epe\ge 2.0\cdot 10^{-3}$.
\label{tab:355}}
\begin{center}
\begin{tabular}{|c|c|c|c|}\hline
  $B^{(3/2)}_8$ & 
  $(\IM\lambda_t)^{\rm min}$ & $Br(K_L\to\pi^0\nu\bar\nu)^{\rm min}$&
$Br(K_L\to\pi^0e^+e^-)_{\rm dir}^{\rm min}$ \\ \hline
0.6 & $ 1.14\cdot 10^{-4}$ &$1.8\cdot 10^{-11}$  &$3.0\cdot 10^{-12}$ \\
0.8 & $ 1.27\cdot 10^{-4}$ &$2.2\cdot 10^{-11}$  & $3.7\cdot 10^{-12}$\\
1.0 & $ 1.42\cdot 10^{-4}$ &$2.7\cdot 10^{-11}$  &$4.7\cdot 10^{-12}$ \\
  \hline
\end{tabular}
\end{center}
\end{table}

\subsection{Impact on $K_L\to\pi^0\nu\bar\nu$ and $K_L\to\pi^0e^+e^-$}
The rare decay $K_L\to\pi^0\nu\bar\nu$ is the cleanest decay in
the field of K-decays. It proceeds almost entirely through
direct CP violation \cite{littenberg:89} and after the inclusion 
of NLO QCD corrections \cite{BB2}
the theoretical uncertainties in the branching ratio are
at the level of $1-2\%$. Similarly the contribution of 
direct CP-violation to the decay $K_L\to\pi^0e^+e^-$ is very
clean. Using the known formulae for these decays \cite{AJBLH,BB2}
and scanning the parameters given in table~\ref{tab:inputparams}
we find:
\be\label{KL0}
1.6\cdot 10^{-11}\le Br(K_L\to\pi^0\nu\bar\nu)\le 3.9\cdot 10^{-11}
\ee
\be\label{KLE}
2.8\cdot 10^{-12}\le Br(K_L\to\pi^0 e^+e^-)_{\rm dir}
\le 6.5\cdot 10^{-12}
\ee
Since these branching ratios are proportional to $(\IM\lambda_t)^2$
any impact of $\epe$ on the latter CKM factor will also 
modify these estimates. We illustrate this in table~\ref{tab:355}
where an improved lower bound on $\IM\lambda_t$ 
implies improved lower bounds on the branching ratios in question.
With decreasing $\bsi$ and increasing $\ms$ these lower bounds
continue to improve excluding a large fraction of the
ranges in (\ref{KL0}) and (\ref{KLE}).
In obtaining the results in table~ \ref{tab:355} correlations
in $\mt$ and the CKM parameters between $\epe$, $\eps$ and
the branching ratios for the decays considered have been taken into
account. 
Unfortunately, due to
 large hadronic uncertainties in $\epe$, no strong conclusions 
can be reached at present.

In the future the situation will be reversed.
As pointed out in \cite{BB96} the cleanest  measurement
of $\IM\lambda_t$ is offered by $Br(\klpn)$:
\begin{equation}\label{imlta}
\IM\lambda_t=1.41\cdot 10^{-4} 
\left[\frac{165\gev}{\mt (\mt)}\right]^{1.15}
\left[\frac{Br(\klpn)}{3\cdot 10^{-11}}\right]^{1/2}\,.
\end{equation}
Once $\IM\lambda_t$ is extracted in this manner it can
be used in $\epe$ thereby somewhat reducing the
uncertainties in the estimate of this ratio. 

\section{Implications for Physics Beyond the Standard Model}
\subsection{General Comments}
We have seen that the Standard Model estimates of $\epe$ are
generally below the experimental results from NA31 and KTeV.
In view of the large theoretical uncertainties it is, however,
impossible
at present to conclude that new physics is signaled by the
$\epe$-data. 
Still, we can make a few general comments on the
extensions of the Standard Model with respect to $\epe$:

\begin{itemize}
\item
In models where the phase of the CKM matrix is the only
source of CP violation, the modifications with respect to
the Standard Model come through new loop contributions to
$\eps$ and $\epe$. If the new contributions to $\eps$ are
positive and the contributions to $F_{\varepsilon'}$ are
negative, then $\IM \lambda_t$, $F_{\varepsilon'}$
and consequently $\epe$ are smaller than in the Standard Model
putting these models into difficulties. An example of
this disfavoured situation is the two-Higgs doublet
model II in which $\epe$ has been analysed a long time ago 
\cite{BBHLS}. We will update this analysis below.
\item
In the Minimal Supersymmetric Standard Model, the
last analysis of $\epe$ after the top quark discovery has
been performed in \cite{GG95}. Here in addition to charged
Higgs exchanges in loop diagrams, also charginos contribute.
The chargino contribution to $\eps$ has always the effect of decreasing
$\IM\lambda_t$. However, depending on the choice of the
supersymmetric parameters, the chargino contribution
to $F_{\varepsilon'}$ can have either sign. Consequently,
$\epe$ in the MSSM can be enhanced with respect to the Standard
Model expectations for a suitable choice of parameters, low
values of chargino (stop) masses and high charged Higgs
masses. Yet, as stressed in \cite{GG95}, generally
$F_{\varepsilon'}$ is further suppressed by chargino
contributions and the most conspicuous effect of
minimal supersymmetry is a depletion of $\epe$.
\item
The situation can be different in more general
models in which there are more parameters than
in the two Higgs doublet model II and in the MSSM, in particular
new CP violating phases.
As an example, in more general supersymmetric models
$\epe$ can be made consistent with experimental
findings \cite{MM99,GMS}. Unfortunately, in view of the large number
of free  parameters
such models are not very predictive.
Similar comments apply to models with 
anomalous gauge couplings \cite{HE} and models with additional
fermions and gauge bosons \cite{Frampton}
in which new positive contributions to $\epe$ are in principle
possible. A recent discussion of new physics effects
in $\epe$ can also be found in \cite{Nierste}.
In the past, there have of course been several other analyses of 
$\epe$
in the extensions of the Standard Model but a review of
these analyses is clearly beyond the scope of this paper.
\item
Finally, models with an enhanced $\bar s d Z$ vertex,
considered in \cite{ISI},  can give rise to large
contributions to $\epe$ as pointed out in \cite{BSII}.
As analyzed in the latter paper, in these models there
exist interesting connections between $\epe$ and rare K decays. 
\end{itemize}
\subsection{An Update on $\epe$ in the Two-Higgs Doublet Model II}
A detailed renormalization group analysis of $\epe$ in
the Two-Higgs Doublet Model II \cite{Abbott} 
 has been presented in \cite{BBHLS}.
It has been found that due to additional {\it positive}
charged Higgs contributions to $\eps$ and corresponding
{\it negative} contributions to $F_{\varepsilon'}$
through the increase of the importance of $Z^0$-penguin
diagrams, the ratio $\epe$ is suppressed with
respect to the Standard Model expectations. Since this
analysis goes back to 1990 and several input parameters,
in particular $\mt$, have been modified we would like to
update this analysis.

We recall that the two new parameters relevant for our
analysis are the charged Higgs mass (${\rm M_H}$) and $\tan\beta$,
the ratio of the two vacuum expectation values.
The expressions for the new contributions with charged Higgs
exchanges to $\eps$ are rather complicated and  will not be
repeated here. They can be found in Section 3 of \cite{BBHLS}.
The QCD corrections to these contributions are given there 
in the leading logarithmic approximation. 
As of 1999 only NLO corrections to box diagram contributions 
with internal top-quark exchanges to $B^0-\bar B^0$ mixing
are known \cite{Dresden}.
Unfortunately the NLO QCD analysis for  $\varepsilon$ in
the 2HDMII is still lacking. For this reason we have
used the leading order expressions for the Higgs
contributions to $\eps$ \cite{BBHLS} except for the
box diagram contributions with internal top-quark exchanges 
where we took the NLO QCD factor obtained in the Standard Model.
While such a treatment is clearly an approximation, 
it is sufficient for our purposes.

The analysis of $F_{\varepsilon'}$  on the other hand can be
done fully at the NLO level. We only have to add to the functions
$X_0(x_t)$, $Y_0(x_t)$, $Z_0(x_t)$ and $E_0(x_t)$ the contributions
from charged Higgs exchanges. They are given as follows:

\be
\Delta X_0=\Delta Y_0 = \frac{x_t}{\tan^2\beta}
\left[\frac{y}{8(y-1)}-\frac{y}{8(x-1)^2}\log{y}\right]
\ee

\be
\Delta Z_0= \Delta X_0 +\frac{1}{4} \frac{1}{\tan^2\beta}D_H(y)
\ee
\be
\Delta E_0 = \frac{1}{\tan^2\beta}
\left[\frac{y\left(7y^2-29y+16\right)}{36\left(y-1\right)^3}+
\frac{y\left(3y-2\right)}{6\left(y-1\right)^4}
\log{y}\right]
\ee
where
\be
D_{H}(y)=\frac{y\left(47y^2-79y+38\right)}{108\left(y-1\right)^3}+
\frac{y\left(-3y^3+6y-4\right)}{18\left(y-1\right)^4}
\log{y}
\ee
with $y=\mt^2/{\rm M_H}^2$.

We observe that all new contributions to $F_{\varepsilon'}$
 are inversely proportional
to $\tan^2 \beta$. In $\eps$ they are inversely
proportional to $\tan^2 \beta$ and $\tan^4 \beta$.
This should be contrasted with the case of $B\to X_s\gamma$
where there are new contributions with charged Higgs
exchanges, which do not involve $\tan\beta$. Thus $\epe$
is more sensitive to $\tan\beta$ than $B\to X_s\gamma$.
This implies that in principle a better constraint for
$\tan\beta$ could be obtained from $\epe$ than from the
latter decay.

\begin{figure}
\begin{center}
% GNUPLOT: LaTeX picture with Postscript
\setlength{\unitlength}{0.1bp}
\begin{picture}(4539,2808)(0,0)
\special{psfile=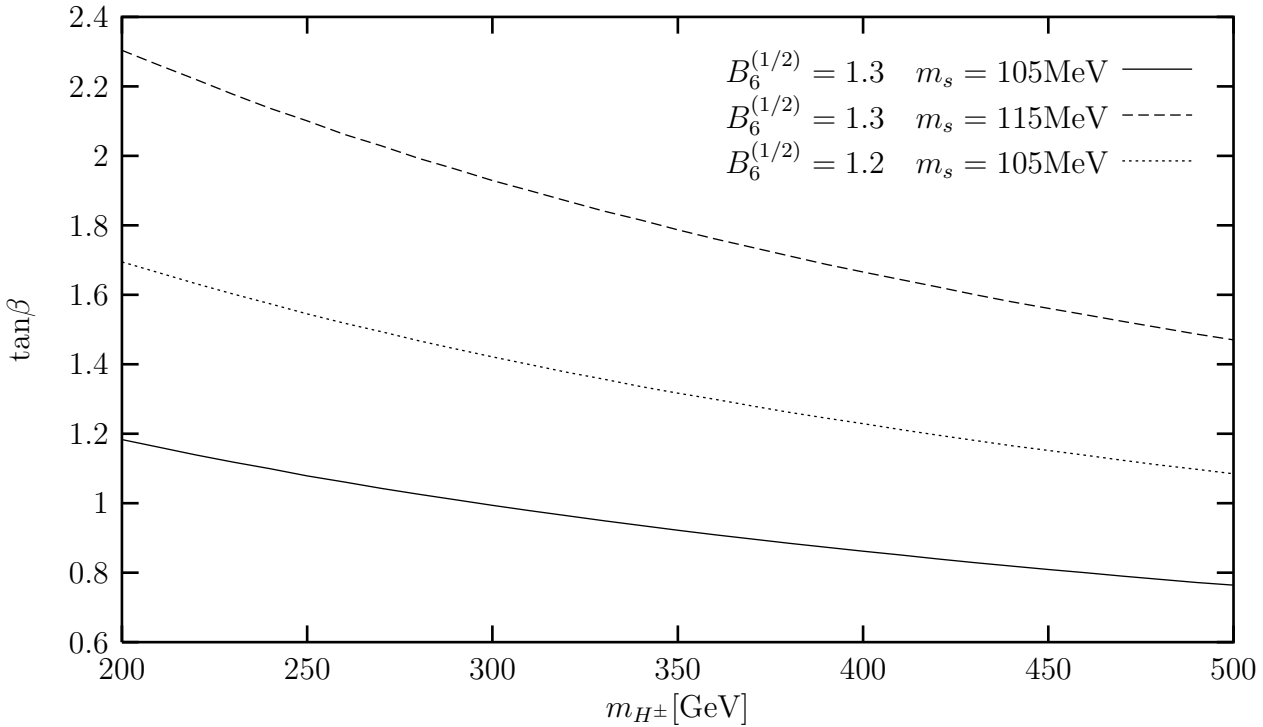 llx=0 lly=0 urx=908 ury=655 rwi=9080}
\put(4070,2165){\makebox(0,0)[r]{$B_6^{(1/2)}=1.2 \quad m_s=105 {\rm MeV}$}}
\put(4070,2338){\makebox(0,0)[r]{$B_6^{(1/2)}=1.3 \quad m_s=115 {\rm MeV}$}}
\put(4070,2511){\makebox(0,0)[r]{$B_6^{(1/2)}=1.3 \quad m_s=105 {\rm MeV}$}}
\put(2444,100){\makebox(0,0){$m_{H^{\pm}} {\rm [GeV]}$}}
\put(0,1529){%
\special{ps: gsave currentpoint currentpoint translate
270 rotate neg exch neg exch translate}%
\makebox(0,0)[b]{\shortstack{${\rm tan \beta}$}}%
\special{ps: currentpoint grestore moveto}%
}
\put(4539,250){\makebox(0,0){500}}
\put(3841,250){\makebox(0,0){450}}
\put(3143,250){\makebox(0,0){400}}
\put(2445,250){\makebox(0,0){350}}
\put(1746,250){\makebox(0,0){300}}
\put(1048,250){\makebox(0,0){250}}
\put(350,250){\makebox(0,0){200}}
\put(300,2708){\makebox(0,0)[r]{2.4}}
\put(300,2446){\makebox(0,0)[r]{2.2}}
\put(300,2184){\makebox(0,0)[r]{2}}
\put(300,1922){\makebox(0,0)[r]{1.8}}
\put(300,1660){\makebox(0,0)[r]{1.6}}
\put(300,1398){\makebox(0,0)[r]{1.4}}
\put(300,1136){\makebox(0,0)[r]{1.2}}
\put(300,874){\makebox(0,0)[r]{1}}
\put(300,612){\makebox(0,0)[r]{0.8}}
\put(300,350){\makebox(0,0)[r]{0.6}}
\end{picture}
\end{center}
\vspace{-6mm}
\caption{Lower bound on $\tan\beta$ as a function of ${\rm M_H}$
 consistent with $\epe\ge 2.0\cdot 10^{-3}$.}
\label{higgs}
\end{figure}

It is obvious from this discussion that $\epe$ in the 2HDMII
is lower than in the Standard Model for any choice of input
parameters. Consequently, for low $\rm{M_H}$ and $\tan\beta$,
the ratio $\epe$ is generally well below the experimental
data. On the other hand if the Standard Model is consistent with
the experimental value of $\epe$, it is possible to put a lower
bound on $\tan\beta$ as a function of $\rm{M_H}$. 
In fig.~\ref{higgs}
we show the result of such an analysis for $\bei=0.8$
and selected values of $\bsi$ and $\ms$.
The remaining parameters have been scanned in the ranges
given in table~\ref{tab:inputparams}. 
We require $\epe\ge 2.0\cdot 10^{-3}$.
We observe that for the
lowest values of ${\rm M_H}\approx 200\gev$
allowed by the $B\to X_s\gamma$ decay \cite{GAMB}-\cite{strum},
$\ms(\mc)=105\mev$ and $\bsi=1.3$ the lower
bound on $\tan\beta$ is similar to the one obtained from
$B\to X_s\gamma$. For higher values of $\ms(\mc)$  and
lower values of $\bsi$ the bound on $\tan\beta$
becomes stronger than from $B\to X_s\gamma$.  
\section{Conclusions and Outlook}
We have presented a new analysis of $\epe$ in the Standard Model in view of 
the recent KTeV measurement of this ratio, which together with 
the previous NA31 
result firmly establishes direct CP violation in nature. 
Compared with our 1996 analysis \cite{BJL96a}, 
the present analysis uses improved 
values of $|V_{ub}|$, $\vcb$, $\mt$, $\Lms^{(4)}$ and $\ms$
 as well as new insights in the hadronic parameters $\bsi$, $\bei$ and
$\hat B_K$. Our findings are as follows:

\begin{itemize}
\item
The estimates of $\epe$ in the Standard Model are typically below 
the experimental data. Our Monte Carlo analysis gives  
\begin{equation}\label{eperangefinal}
\epe =\left\{ \begin{array}{ll}
( 7.7^{~+6.0}_{~-3.5}) \cdot 10^{-4} & {\rm (NDR)} \\
( 5.2^{~+4.6}_{~-2.7}) \cdot 10^{-4} & {\rm (HV)} \end{array} \right.
\end{equation}
The difference between these two results indicates the left over
renormalization scheme dependence.

\item
On the other hand a simple scanning of all input parameters  
gives
\begin{equation}
1.05 \cdot 10^{-4} \le \epe \le 28.8 \cdot 10^{-4}\qquad {\rm (NDR)}
\label{eper}
\end{equation}
and
\begin{equation}
0.26 \cdot 10^{-4} \le \epe \le 22.0 \cdot 10^{-4}\qquad {\rm (HV)}
\label{eperh}
\end{equation}
This means that for suitably chosen parameters, $\epe$ in the Standard 
Model can be made consistent with data. However, this happens only if all 
relevant parameters are simultaneously close to their extreme values. 
This is clearly seen in table~\ref{tab:31731} and fig.~\ref{g2}.
Moreover, the probability density distributions for $\epe$ in
fig.~\ref{g1} indicates that values of $\epe$ in the ball park of
NA31 and KTeV results are rather improbable.
\item
Unfortunately, in view of very large hadronic and substantial parametric 
uncertainties, it 
is impossible to conclude at present whether new physics contributions are 
indeed required to fit the data. 
Similarly it is difficult to conclude what is precisely the impact of 
the $\epe$-data on the CKM matrix. However, there are indications as seen 
in table~\ref{tab:355} that the lower limit on 
$\IM\lambda_t$ is improved. The same applies to the lower limits for the 
branching ratios for $K_L\to\pi^0\nu\bar\nu$ and 
$K_L\to\pi^0 e^+ e^-$ decays.
\item
Finally, we have pointed out that the $\epe$ data puts models in which there 
are new positive contributions to $\eps$ and negative contibutions to 
$\varepsilon'$ in serious difficulties. In particular 
we have analyzed $\epe$ in the 2HDMII demonstrating that with improved 
hadronic matrix elements this model can either be ruled out or a  
powerful lower bound on $\tan\beta$ can 
be obtained from $\epe$.
\end{itemize}

The fact that one cannot firmly conclude at present that the data
for $\epe$ requires new physics is rather unfortunate.
In an analogous situation in the very clean rare decays
$K\to\pi\nu\bar\nu$ a departure of the experimental
result from the Standard Model expectations by only
$30\%$ would give a clear signal for new physics.
This will indeed be the case if the
improved measurements of the $Br(K^+\to\pi^+\nu\bar\nu)$
from BNL787 collaboration at Brookhaven \cite{Adler}
find this branching ratio above $1.5 \cdot 10^{-10}$.
All efforts should be made to measure this branching
ratio and the branching ratio for $K_L\to\pi^0\nu\bar\nu$,
which while being directly CP violating is almost
free of theoretical uncertainties.

The future of $\epe$ in the Standard Model and in its extensions depends on 
the progress in the reduction of parametric and hadronic uncertainties. 
We have analyzed these uncertainties in 
detail in Section 3 with the results given in table~\ref{tab:3}.

Concerning parametric uncertainties related to 
$|V_{ub}|$, $\vcb$, $\mt$ and $\Lms^{(4)}$,
we expect that they should be reduced considerably in the coming years. 
This will, however, result only in a modest reduction of the 
total uncertainty in $\epe$. In this respect a measurement of 
$\IM\lambda_t$ in a very clean decay like $K_L\to\pi^0\nu\bar\nu$ would 
be very useful. 

A real progress in estimating $\epe$ will only be made if the 
non-perturbative parameters $\hat B_K$, $\bsi$, $\bei$ and $\OEE$ as 
well as the strange quark mass $\ms$ will be brought under control. 
In particular the sensitivity of non-perturbative methods to
$\mu$ and renormalization scheme dependences of $\bsi$ and $\bei$
is clearly desirable. 
We expect that considerable 
progress on $\hat B_K$ and $\bei$ should be made in the coming years 
through improved lattice calculations. 
Progress on $\OEE$ should also be possible in the near future.
Moreover, as various estimates of $\hat B_K$, $\bei$ and $\OEE$ 
by means of
several non-perturbative methods are compatible with each other
we do not expect big surprises here. Similar comments
apply to $|V_{ub}|$, $\vcb$, $\mt$ and $\Lms^{(4)}$.

On the other hand, it appears that it will take 
longer to obtain acceptable values for $\ms$ and $\bsi$.
In view of the bounds \cite{DERAF}-\cite{Dosch}, it is difficult
to imagine that $\ms(\mc)\le 105\mev$. Consequently we
expect that future improved estimates of $\ms$ will most probably
exclude the lowest values of $\ms$ considered in this paper.
This would simultaneously exclude the highest values for $\epe$ 
obtained
by us unless $\bsi$ is found to be higher than used here.
In this respect improved estimates of $\bsi$, if found 
substantially higher than unity, could have considerable
impact on our analysis. Finally, it should be stressed that
future lattice calculations will give the full matrix
elements without the necessity to use separately
$(\bsi,\bei)$ and $\ms$.

In any case $\epe$ already played a decisive role in establishing direct 
CP violation in nature and its rather large value gives additional strong 
motivation for searching for this phenomenon in 
cleaner K decays like 
$K_L\to\pi^0\nu\bar\nu$ and 
$K_L\to\pi^0 e^+ e^-$, in B decays, in D decays and elsewhere.

{\bf Acknowledgements}\\
We would like to thank S. Bertolini, G. Buchalla, M. Ciuchini,
E. Franco, P. Gambino, 
T. Hambye, L. Littenberg, G. Martinelli, P. Soldan and J. Urban
for discussions.

This work has been supported by the German Bundesministerium f\"ur 
Bildung und Forschung
under contract 06 TM 874 and DFG Project Li 519/2-2.

\vfill\eject

\end{document}